\documentclass{aa}  

\usepackage{graphicx}
\usepackage{longtable}

\usepackage{lscape}
\usepackage{eqnarray,amsmath}
\usepackage{txfonts}
\usepackage{xcolor}
\begin{document}

\title{Magnetically arrested accretion disks launching structured jets in application to GRB and AGN engines
  %\LEt{ AA avoids direct questions in the title and text. If this title isn't suitable, please suggest another phrasing.}
}
\titlerunning{Magnetized accretion and structured jets}
  
\author{Agnieszka Janiuk
          \inst{1}\fnmsep\thanks{agnes@cft.edu.pl}
          \and
          Bestin James \inst{1}
}
%\email{agnieszka.janiuk@gmail.com}
%\author{Agnieszka Janiuk}
\institute{Center for Theoretical Physics, Polish Academy of Sciences, 
Al. Lotnikow 32/46, 
02-668 Warsaw, Poland \\
}
%\email{agnes@cft.edu.pl}
%\author{Bestin James}
%\affil{Center for Theoretical Physics, Polish Academy of Sciences, Al. Lotnikow 32/46, 
%02-668 Warsaw, Poland}
\newcommand{\lrz}{\gamma}
\newcommand{\enth}{\xi}
\newcommand{\hslope}{h}
\newcommand{\polind}{\hat \Gamma}
%\begin{abstract}
    \abstract
  {We explore the formation, energetics, and geometry of relativistic jets along with the  variability of their central engine.
 We study both fast and slowly rotating black holes and address our simulations to
 active galaxy centers
as well as gamma ray burst engines. }
{The structured jets are postulated to account for emission properties of high energy sources across the mass scale, launched from stellar mass black holes in gamma ray bursts (GRBs) and from supermassive black holes
in  active galactic nuclei (AGNs).
Their active cores contain magnetized accretion disks and the rotation of the Kerr black hole provides a mechanism for launching relativistic jets.
This process works most effectively if the mode of accretion turns out to be
magnetically arrested. In this mode, the modulation of jets launched from the engine is related
to internal instabilities in the accretion flow that operate on smallest time and spatial scales.
As these scales are related to the light-crossing time and the
black hole gravitational radius, the universal model of jet-disk connection
is expected to scale with the mass of the black hole.}
{We investigated the jet-disk connection by means of 3D general relativistic magneto-hydrodynamical simulations of the
magnetically arrested disk accretion in Kerr geometry. We also quantified the variability of the disk by means of a Fourier analysis.}
{We found that the system evolution is governed by the physical parameters of the engine, such as the black hole spin and disk size, as well as disk magnetization, and we applied our scenarios to typical types of sources
  in AGN and GRB classes. We found that the magnetically arrested disk (MAD) scenario is applicable to AGN engines and supports persistent
  jet emissions. It can also be applied to GRBs, as it gives the variability pattern roughly consistent with observations. However, in some cases, strong magnetic fields may lead 
to jet quenching, and this effect is found to be important mainly for GRB jets. We speculate that it may be related to the strength of magnetically driven winds from the GRB engines.}
{}
%\end{abstract}

%\end{keywords}
   \maketitle
%
%-------------------------------------------------------------------

\section{Introduction}

The central engine of an active galactic nucleus (AGN) is powered by accretion flow
mediated by magnetic fields and centrifugal forces, where the matter accretes onto a rotating or non-rotating black hole \citep{Krolik1999}. 
Accretion powers the relativistic jets of AGN, which manifest their complex structure on pc, kpc, and Mpc scales \citep{Harris2006}.
The power of these jets, in many radio loud galaxies, is comparable to their accretion rates, $P_{jet}\sim \dot M c^{2}$ \citep{Punsly2007, Sikora2013}. It has been argued that for this level of efficiency among jets produced by the Blandford-Znajek mechanism, \citep{Blandoford_Znajek_1977}, the magnetic flux accumulated on the black hole horizon must be extremely high \citep{SikoraBegelman2013}, so that the accretion proceeds in the magnetically dominated (MAD) mode \citep{Igumenshchev2008, Tchekhovskoy_2011, McKinney_2012, 2012MNRAS.426.3241N, Tchekhovskoy_Giannios2015}.
A similar engine (on scales of one order of magnitude smaller in size) operates in gamma ray bursts (GRBs), where it launches baryon-free, relativistically expanding, and collimated jets \citep{Sarietal1999, Rhoads1999}.

An open question abounds regarding the main physical parameter responsible for the jets efficiency and the driver of the AGN radio-loudness \citep{Rusinek2020}. 
Previous works (e.g. \cite{moderski}) have suggested that
the black hole spin is an important quantity in this context. However, more recent works (e.g., \cite{Falcke2013}) have claimed that the spin is not the main driver and contributes only to the scatter of radio-loudness distribution of active galaxies.
As  found by \cite{Avara2016} %\LEt{ If the reference is part of the sentence, only the year should be in parentheses.}
in their gamma-ray magnetohydrodynamic (GR MHD) simulations, the jets launched from moderately spinning black holes, $a=0.5$, are twice as efficient for the geometrically thin disks than for thick ones.  However, the radiation-transport simulations presented by
\cite{2018MNRAS.480.3547M} have shown that the radiative efficiency of luminous disks is in line with the
Novikov-Thorne model prediction, while it is unclear how much of the radiation and thermal energy trapped in the outflows could ultimately escape. This effect should be confronted with
the spectral state transitions in stellar black hole binaries and their jets being quenched in their soft states \citep{Fender2004}.

The significant part of the jet acceleration after it leaves the black hole region is inside the hot cocoon, whose presence may alter the internal energetics (e.g., \citet{Lazzati2017}). Such a cocoon may plausibly be formed from an uncollimated wind outflow that was launched from the accretion disk and driven, for instance, by magnetic fields \citep{Murguia_Berthier2021}.
On the other hand, the imprint of the MRI within the accretion disk at the base of the jet will manifest itself in the variability timescales \citep{Sap2019ApJ}.
Hence, the interplay between this variability and the ultimate jet Lorentz factor should reveal the properties of the magnetic field
configuration preserved in the accretion torus.
From an observational perspective, this variability results from  the
periodic structures seen in the radio brightness of some large-scale jets \citep{Godfrey2012} or in the minute-timescales variability seen in mid-infrared in black hole X-ray binaries \citep{Baglio2018}. Central engine properties can be also imprinted on the observed variability of GRB jets, including correlations between the peak energy and power density spectral slope \citep{Dichiara2016}.

The maximum jet power in the MAD state was estimated to be on the order of betwen three and four times the accretion power, that is, 
$P_{MAD}=1.3 h_{0.3}a^{2}{\dot M} c^{2}$ \citep{DaviesRev}. 
In terms of the magnetic pressure driven onto the black hole,  which is proportional to the square of the magnetic flux, $\Phi^{2}_{BH}$, the balance of the ram pressure of matter, given by the accretion rate, should be satisfied. The relative strength of the ram and magnetic pressure scales, thus, the disk-jet connection, while the dimensionless quantity, $\phi_{BH}=\Phi_{BH}/{\sqrt{\dot M r_{g} c^{2}}}$, defined at the black hole horizon, gives the characteristic threshold for magnetically arrested state. In this state, the dimensionless magnetic flux $\phi_{BH}$ can be in the range between 10-20, which means that it can be stronger than the gravitational force pulling the matter to the black hole and can actively push it back from the horizon. While the material gets accreted, the magnetic flux cannot escape due to the inward pressure. The process is regulated via interchange instabilities and magnetic field reconnections. The detailed picture of this MAD state is described by strong non-linear interactions between magnetic field and the gas, and comprises  the subject of complex numerical simulations.

Gamma ray bursts are the transient sources resulting from a collapse of massive stars, or compact binary mergers, where the relativistic jets launched from the central engine are responsible for the observed emission (see reviews in e.g.,  \cite{Lee2007,Corsi2021}). The prompt emission in gamma rays is modulated on very short timescales, which can reflect the variable activity of the central engine \citep{Kobayashi1997}. In the MAD state, this variability is governed by the free-fall timescale of matter in the region where it is stalled by magnetic fields. The parametric model proposed by \cite{LloydRonning2018} relates the degree of arrestedness (the so-called "MAD-ness parameter"), the size of arrested region, and the mass of black hole with the observable variability timescales -- which are on the order of $\sim 1$ second or less, for typical parameters. Observed correlations between the minimum timescale of variability and the jet Lorentz factor \citep{wu2016}, and between the computed power-density slope and the black hole spin \citep{Janiuk2021} seem to confirm the crucial role of the central engine in governing this variability.

In this work, we perform a set of numerical simulations of the 
central engine as composed of the 
rotating black hole surrounded by accretion torus.
Our models can be applicable to both AGN cores and GRB centers.
 We address here the widely-adopted MAD paradigm, which has been aimed at explaining the power released in the jets launched from the magnetically arrested disks and their time variability. The novelty of our approach allows us to infer the
  observationally testable results for a wide range of sources, which scale from stellar mass black holes to the supermassive ones. In contrast to the previous studies which were focused, for instance, on the specific aspects of the state transitions in black-hole X-ray binaries \citep{dexter} or the changing-look AGNs \citep{2021scepi},
here we model the short-term variability of the accreting black holes across their mass scale.

Our numerical setup is based on the full general relativistic MHD framework and the simulation results are scalable with the black hole mass over many orders of magnitude.
We show fully non-axisymmetric 3D simulation results
in the fixed Kerr background metric that are sufficient for a negligible mass
and spin change of the black hole (for other treatment, see e.g., \citet{Janiuk2018}).
The torus matter is highly magnetized and its 
accurate evolution is followed when the MHD turbulence
\citep{Balbus_Hawley_1991} drives the mass inflow.
In addition, the magnetic field is responsible for launching 
the uncollimated outflows of the plasma, in the form of magnetically driven winds, while the rotation of the black hole
powers the bipolar jets.

The numerical scheme is based on the
GRMHD code HARM \citep{Gammie_2003, Noble_et_all_2006}. The code has been extended to 3D and efficiently parallelized with a hybrid MPI-OpenMP technique \citep{JSFI177}. Our code is publicly available under the name HARM-COOL \footnote{https://github.com/agnieszkajaniuk/HARM\_COOL},
%\footnote{https://www.iau.org/science/scientific-bodies/commissions/B1/info/useful\_resources/},
as it contains the module for neutrino cooling of dense and hot plasma (this specific module currently  only operates in 2D).
The current paper is a follow-up of our previous studies \citep{Sap2019ApJ}, which presented the axisymmetric model. For efficient 3D runs, we adopted a technique of minimizing the total entropy in the cells where the magnetic pressure is extremely high.
The initial setup of our code is also changed due to the breaking of axial symmetry in the flow
via random perturbations of internal energy. In this way, we allow the plasma to start leaking onto the black hole through the finger-like structures, within the magnetically arrested regions. As a result, we are able to probe much more highly magnetized configurations.

We study models with several values of the black hole spin parameter, broadly ranging from fast to non-rotating black holes.
In addition, we probe two limits for the accretion disk size, as determined by the initial radius of the pressure maximum.
The study is addressed to both stellar mass black hole-jet systems in the GRB engines and to the AGN engines.

This article is organized as follows. In Section 2, we present the simulation scheme and the initial configuration of our model.
 The results showing the general properties of the flow and the
  results of the jet power and efficiency analysis are presented in Section 3.
 Our discussion and conclusions make up Section 4.

\section{The simulation setup}

We use the general relativistic magneto-hydrodynamic code, \textit{HARM} \citep{Gammie_2003, Noble_et_all_2006, Sap2019ApJ}, with a fixed background
Kerr metric, that is, we neglected the effects of self gravity and the BH spin changes (cf. \citet{Janiuk2018}).
The \textit{HARM} code is a finite-volume, shock-capturing scheme to solve 
the hyperbolic system of the partial differential equations of GR MHD.
The numerical scheme is based on GR MHD equations where
the energy-momentum tensor, $T^{\mu\nu}$, is contributed by the gas and electromagnetic fields:
\begin{eqnarray*}
{T_{\left(m\right)}}^{\mu\nu}= \rho \enth u^\mu u^\nu + p g^{\mu\nu}, \\
{T_{\left(em\right)}}^{\mu\nu}=b^\kappa b_\kappa u^\mu u^\nu+\frac{1}{2} b^\kappa b_\kappa g^{\mu\nu} - b^\mu b^\nu,\\
T^{\mu\nu}={T_{\left(m\right)}}^{\mu\nu}+{T_{\left(em\right)}}^{\mu\nu}.
\end{eqnarray*}
Here $u^{\mu}$ is the four-velocity of gas, $u$ denotes internal energy density,  $b^{\mu}$ is the magnetic four-vector %\LEt{ is there a word missing after four-vector?}
 and
$\enth$ is the fluid specific enthalpy.
The continuity and momentum conservation equation is expressed as:
\begin{equation}
\
(\rho u^{\mu})_{;\mu} = 0,
\hspace{1cm}
T^{\mu}_{\nu;\mu} = 0.
\end{equation}
These are brought into a conservative form by implementing a Harten, Lax, and van Leer (HLL) solver to numerically calculate  the corresponding fluxes. 
In terms of the Boyer-Lindquist coordinates, $\left(r,\theta,\phi\right)$, the black hole is located at $0<r \leq r_{\rm h}$, where $r_{\rm h} = \left(1+\sqrt{1-a^{2}}\right) r_{\rm g}$ is the horizon radius of a rotating black hole with mass, $M,$ and angular momentum, $J,$ in geometrized units, $r_{\rm g}=G M /c^2$, and $a$ is the dimensionless Kerr parameter, $a=J/(Mc), 0 \leq a \leq 1$. 
In our simulations, we adopt the rotating black hole, with $a=0.9$.

The {\it HARM} code doesn't perform the integration in the Boyer-Lindquist coordinates, but instead in the so called Modified Kerr-Schild ones: $\left(t,x^{(1)},x^{(2)},\phi\right)$ \citep{Noble_et_all_2006}. The transformation between the coordinate systems is given by:

\begin{eqnarray*}
r= R_0 + \exp\left[{x^{(1)}}\right]
\\
\theta = \frac{\pi}{2} \left(1 + x^{(2)}\right) + \frac{1 - \hslope}{2} \sin\left[\pi\left(1 +x^{(2)}\right)\right]
\end{eqnarray*}
where $R_0$ is the innermost radial distance of the grid, $0 \leq x^{(2)} \leq 1$, and $\hslope$ is a parameter that determines the concentration of points at the mid-plane. In our models, we use $h=0.3$ (noting that for $h=1$ and a uniform grid on $x^{(2)}$, we obtain an equally spaced grid on $\theta$, while for $\hslope=1$ the points concentrate on the mid plane). The exponential resolution in the $r$-direction leads to higher resolution and it is adjusted to resolve the initial propagation of the outflow.

In the current simulation, we use additional modifications of the original grid, to optimally resolve  the regions closest to the black hole horizon and the polar axis.
The cylindrified coordinates are used inside the radius $R_{\rm cyl}=3.5$ and inside the angle $\theta_{\rm cyl}=-1+1/N2$.
This restricts grid reshaping
to the one
single grid cell closest to the pole
and minimizes the resulting change in resolution at the poles.

Our computational domain is spanning $10^{5}$ gravitational radii, while the grid
  cells are scaled with distance from the center.
The inner and outer radius are set to $R_{\rm in} = 0.87\times(1. + \sqrt(1. - a^{2}))$, and $R_{\rm out} = 10^{5}$.
The progressively sparser grid starts at $R_{\rm br}=400$,
with $x^{(1)}_{i}=x^{(1)}_{i-1}+1/C \times \delta x$,
where
$\delta x = x^{(1)}_{i} - (\log(R_{br} -R_{0}))^{4}$. We use $C=1$ to limit the hyper-exponentiation.
Our grid resolution is $N1 \times N2 \times N3 = 288 \times 256 \times 256$ in the $r, \theta$, and $\phi$ directions.

The boundary conditions implemented in the code are free outflow boundary in the radial direction, transmitting and reflecting in the polar direction, and periodic in the azimuthal direction. In the radial direction, primitive variables (density, internal energy, and radial magnetic field component), are copied from the boundary zones to the two ghost zones. The radial velocity and the polar and azimuthal components of the magnetic field are also extrapolated.
In the polar direction, the primitive variables are copied to the ghost zones and we have two such zones, located on the opposite side of the polar axis.
  The $\theta$-components of velocity and magnetic field change signs at the polar axis and, in addition, the transverse velocity at the polar axis is linearly interpolated over two innermost cells in $\theta$.

\subsection{The torus initial configuration}

\begin{figure}
\centering
    \includegraphics[width=0.45\textwidth]{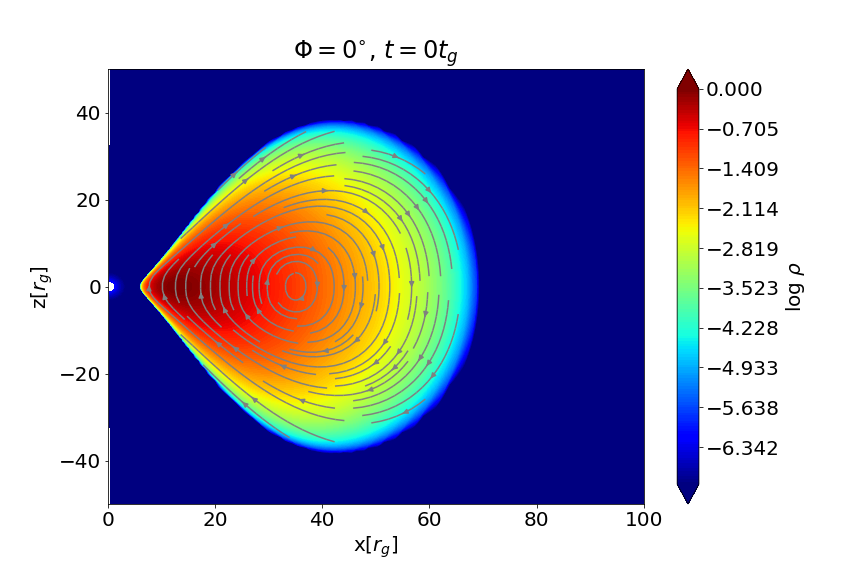}
    \includegraphics[width=0.45\textwidth]{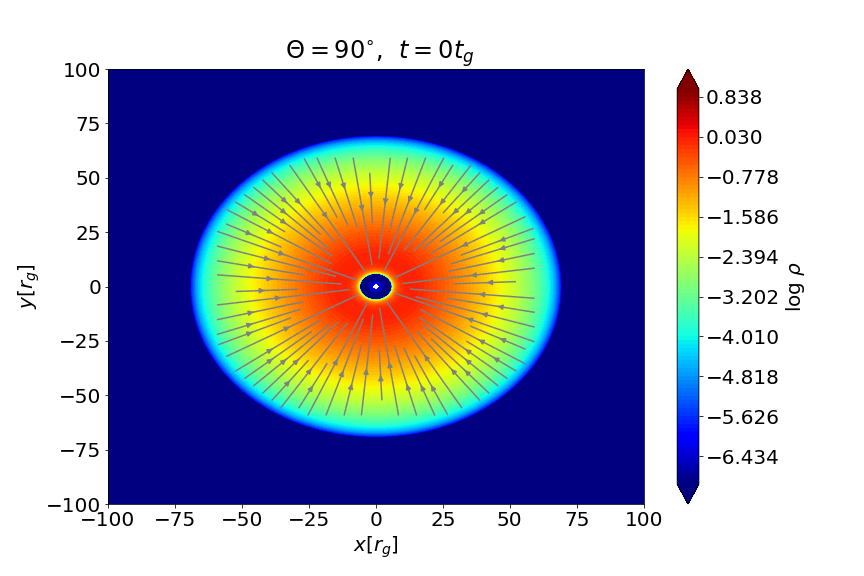} 
    \caption{Initial state of the model GRB-HS100, in the polar and equatorial cuts, as shown in the top and bottom panels, respectively. Maps present the density distribution in the FM torus configuration and poloidal magnetic field.}
    \label{fig:initial}
\end{figure}

The  accreting material is modeled following \citet{Fishbone_Moncrief_1976_ApJ} (hereafter, FM) based on the analytic solution of a constant specific 
angular momentum, in a steady-state configuration of a pressure-supported ideal fluid in the Kerr black hole potential.
The initial configuration of our fiducial model is shown in Figure \ref{fig:initial}.

The location of the material reservoir is determined by the radial distance of the innermost cusp of the FM torus, $r_{\rm in}$, and the distance where the maximum pressure occurs, $r_{\rm max}$.
In the current simulations, we adopt values 6 (12) and 13 (25), for  $r_{\rm in}$ and $r_{\rm max}$, respectively.
The relative difference of the two radii determines also the dimension of the torus.
Subsequently, we obtain the angular momentum value and the distribution of the angular velocity along the torus.
For our parameters, and back hole spin of $a=0.9$, it is equal to $l_{\rm spec} = 4.397$, and $l_{spec}=5.556$, in a smaller and larger torus, respectively.
We also note that the torus size and the value of specific angular momentum in the FM solution depend on the Kerr parameter. For small black hole spin, the torus size decreases,
while the specific angular momentum is increased.

The initial torus is embedded in a poloidal magnetic field, prescribed with the vector potential, 
\begin{equation}
A_{\varphi}=\max (({\bar{\rho} \over \rho_{\rm max}} - \rho_{0}) \times r^{5}, 0),
\end{equation}
where $\bar{\rho}$ is the density in the torus (averaged over 2 neighboring cells) and 
$\rho_{\rm max}$ is the density maximum, and we use $\rho_{0}=0.2$, to restrict the initial field to the regions where density in the torus is larger than $0.2 \rho_{\max}$.
The factor of $r^{5}$ ensures that higher magnetic flux will be brought
  onto the black hole horizon from larger distances, as the evolution proceeds.
  Other spatial components of the initial vector potential are zero, $A_{r}=A_{\theta}=0$; hence, the magnetic field vector will have only $B_{r}$ and $B_{\theta}$ components
  The vector potential is related to the Faraday tensor, $F_{\mu \nu}=\partial_{\mu}A_{\mu}-\partial_{\nu}A_{\mu}$, and we thus obtain the magnetic field
  four-vectors with $b^{\mu} =-^{*}F^{\mu\nu} u_{\nu}$. It is related to the normal observers magnetic field vector, as $B^{\mu}=\alpha ^{*}F^{\mu t}$, where $\alpha = 1/g_{tt}$ is the lapse (noting that there is no time-component for $B^{\mu}$).

The magnetized flows are, by definition, not in equilibrium and the angular momentum is transported. 
The plasma $\beta-$parameter is
defined as the ratio of the fluid's thermal to the magnetic pressure, $\beta \equiv p_{\rm g} / p_{\rm mag}$.
We normalize the magnetic field in the FM torus to have
$\beta=(\gamma - 1.)*u_{max}/(0.5*b^{2}_{max}) $, where $u_{max}$ is the internal energy at the torus pressure maximum radius. We 
examined several models with different minimal initial $\beta$ of the plasma
(with tested values of 30, 50, and 100).

For the thermal pressure, we adopt the adiabatic equation of state, of the form $p=(\gamma-1) u$, and we use
  the adiabatic index of $\gamma=4/3$.
  Finally, in order to introduce a non-axisymmetric perturbation and allow for the 
  generation of azimuthal modes, permitting the gas to overpass the magnetic barrier, we impose an initial perturbation of internal energy.
  It is given by $u = u_{0} (0.95+0.1C)$, where $C$ is a random number generated in the range (0;1). This perturbation on the order of less than 5\% is typically used in other 3D simulations of weakly magnetized flows (see \citet{Mizuta2018}).

The code works in dimensionless units of $c=G=M=1$. For the purpose of these computations, we keep the scaling, and the black hole mass $M$ does not change during the simulation.
In physical scales, the GRB engines are modeled with the following global parameters: mass of the black hole, $M_{\rm BH}$; the torus mass; and black hole dimensionless spin, $a$.
Typical short GRB parameters are $M_{\rm BH}=3 \,M_{\odot}$, and $a=0.6-0.9$, while the torus mass is about 0.1 $M_{\odot}$.
The latter depends on the configuration of our torus (i.e., the inner and pressure maximum radii) and can be scaled to physical units by assuming a density normalization.
For the GRB case, we adopt the density scaling to keep the mass of torus enclosed in the volume, so that 
  $D_{unit}= M_{unit}/L_{unit}^{3}= 3.4\times 10^{11}$ g cm$^{-3}$.
The magnetic flux is then normalized with $\Phi_{unit}= \sqrt{4\pi D_{unit}c^{2}} L_{unit}^{2}$ (see Table \ref{tab:conversion}).

\begin{table*}[htb]
\centering
\begin{tabular}{lcccccccc}
  %\toprule
  \hline
  \hline
Source & $M_{BH}$ & $M_{disk}^{unit}$ & Time$^{unit}$ & $\dot M^{unit}$ & $\dot E^{unit}$ & $D^{unit}$ & B$^{unit}$ & $\Phi_{B}^{unit}$ \\

       & ($M_{\odot}$)&  ($M_{\odot}$) &             &                & (erg s$^{-1}$) &(g cm$^{-3}$)&  (G)        &                  \\
\hline
\hline
 Short GRB & 3  & 1.5$\cdot 10^{-5}$ & 1.5$\times 10^{-5}$ s & 1 $M_{\odot}$s$^{-1}$ & 1.8 $\cdot 10^{54}$ & 3.4 $\cdot 10^{11}$ & 6.2 $\cdot 10^{16}$ & $1.2 \cdot 10^{28}$ G cm$^{2}$ \\
 Long GRB  & 10 & 1.5$\cdot 10^{-4}$ & 4.9$\times 10^{-5}$ s & 3 $M_{\odot}$s$^{-1}$ & 5.5 $\cdot 10^{54}$ & 9.5 $\cdot 10^{10}$ & 3.2 $\cdot 10^{16}$ & $7.0 \cdot 10^{28}$ G cm$^{2}$ \\

 Sgr A$\star$  &  $ 4 \cdot 10^{6}$ & 6.2$\cdot10^{-10}$ & 6.2$\times 10^{-7}$ yr &  10$^{-3}$ $M_{\odot}$yr$^{-1}$ & 5.6 $\cdot 10^{43}$ & 6.0 $\cdot 10^{-12}$ & 2.6 $\cdot 10^{5}$ & $9.0 \cdot 10^{-8}$ G pc$^{2}$ \\
 M 87          & $ 5 \cdot 10^{9}$ & 7.8$\cdot 10^{-7}$   & 7.8$\times 10^{-4}$ yr &  10$^{-3}$ $M_{\odot}$yr$^{-1}$ & 5.6 $\cdot 10^{43}$& 3.8 $\cdot 10^{-18}$ & 2.6 $\cdot 10^{2}$ & $1.2 \cdot 10^{-5}$ G pc$^{2}$\\
 \hline
 
\end{tabular}
\caption{Conversion units for various astrophysical sources}
\label{tab:conversion}
\end{table*}

    In case of AGN, we can apply our models to low-luminosity AGN, such as Galaxy center SgrA$\star$ or M87. In these sources, the accretion rates are very small and our radiatively inefficient models are still a viable approximation. We therefore adopted two similar scalings for accretion rates, on the order of $10^{-5}$ $M_{\odot}$yr$^{-1}$, for these AGN \citep{1999ApJ...517L.101Q,2005MNRAS.360L..55C,2016MNRAS.457.3801P}. We adopted the black hole mass of $4\times 10^{6} M_{\odot}$ and  $5\times 10^{9} M_{\odot}$, for  SgrA$\star$ and M87, respectively, and we derive our length units, according to their gravitational radius.
We apply the scaling of SgrA$\star$ to the simulation with a non-spinning black hole \citep{2001ApJ...554L..37M, 2011ApJ...735..110B}, while the model with black hole spin of $a=0.9$ represents the M87 case \citep{2020MNRAS.492L..22T}.  
  The magnetic flux on the black hole horizon is expected to be on the order
  of $10^{-10}-0.01$ G pc$^{2}$ \citep{McKinney_2012}.
  
  Density scaling in the accretion disk of AGNs is more arbitrary than for GRBs, as the spacial extension of the disk is much larger than in case of compact binaries and the density drops by many orders of magnitude, from the black hole up to the broad line region \citep{Czerny2016}. At the outskirts of the disk, it can be as low as the insterstellar medium (ISM) density, $\rho_{\rm ISM}\sim 10^{-27}$ g cm$^{-3}$. Also, the disk thickness in AGN is much smaller than its radius, while the self-gravity effects may further reduce the surface density. In all our simulations, the total mass of the disk enclosed in our computational domain, $R_{out}=10^{5} r_{g}$, will be a fraction of the solar mass.
  Magnetic flux is again normalized to physical units with the scaling of density, where  now we adopt
 $D_{unit}=6.03\times 10^{-12}$ g cm$^{-3}$ and $\Phi_{unit}=9.08\times 10^{28}$ G cm$^{2}$ and then $D_{unit}=8.85\times 10^{-18}$ g cm$^{-3}$ and $\Phi_{unit}=1.15\times 10^{32}$ G cm$^{2}$, for  SgrA$\star$ and M87, respectively.

\section{Results}

\begin{figure}
    \centering
    \includegraphics[width=0.4\textwidth]{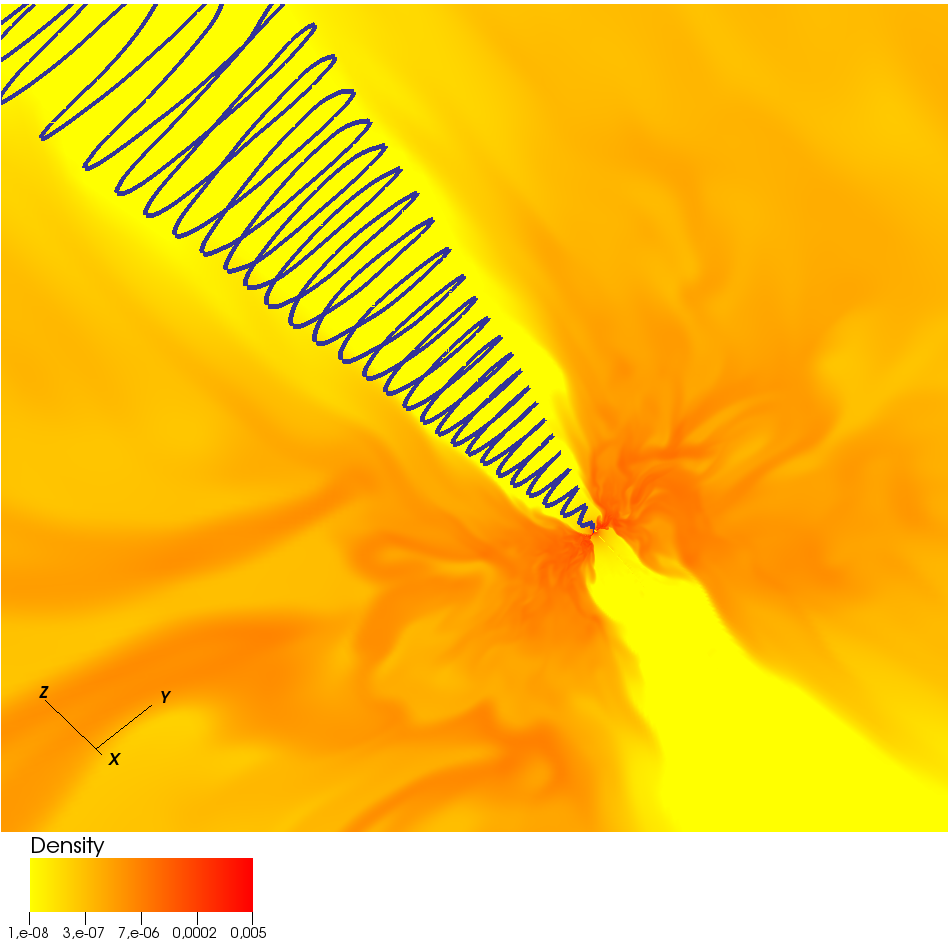}
    \caption{3D view of the inner parts of torus in the end of simulation with $\beta=100$ and black hole spin of $a=0.9$ (model GRB-HS100). The color scale presents density of the flow, while the thick blue line is a visualization of the magnetic field structure within the jet, with a chosen field line originating in the region close to black hole horizon (sphere of radius $r=2$). We note that the plot does not show the full computational domain, but it is zoomed to the innermost $\sim 100$ $r_{g}$. For a GRB model with a black hole of three solar masses in the engine, the physical size in this plot is about $4.5\times 10^{7}$ cm.
     }
    \label{fig:map_Bspiral}
    \end{figure}

 The bipolar Poynting-dominated jets emerge in our GRB and most of AGN models.
  They are collimated close to the black hole by magnetic stress, created
  by rotation of the disk and associated with a disk wind \citep{GlobusLevinson2016}, as well as
  supported by toroidal magnetic field \citep{LevinsonGlobus2017}.
Figure \ref{fig:map_Bspiral} shows an example of such a structure, where the central engine is at the base of the jet.

 \begin{table*}[]
   \centering
   \begin{tabular}{l|c|c|c|c|c|c|c|c|c|r}
     \hline
     \hline
%\title{Summary of the models}

Model & $M_{torus}$ (t=0) & $M_{torus}$ (t=$t_{end}$) & BH spin &  $r_{\rm in}$ & $r_{\rm max}$ & $\beta_{0}$ &  $P_{BZ}$ &  $\langle \dot M_{in} \rangle$ & $\langle \Phi_{B} \rangle$ & $\Delta M_{out}$ \\
\hline
\hline
      GRB-HS100 & 26443  & 12500 & 0.9 & 6.0 & 13.0 & 100 & 0.0102 & 0.147 & 2.121 & 6.464 \\ % /HARM3D/MAD/harm_mad482/exe/Model1a_3D
      GRB-HS50 & 26466  &  12189 & 0.9 & 6.0 & 13.0 & 50 &  0.0013 & 0.143 & 1.338 & 12.89 \\ %HARM3D/MAD/harm_mad482/exe/Model5a_3D
      GRB-HS30  & 26498  & 15634 & 0.9 & 6.0 & 13.0 & 30 &   - &  0.269 & 1.317 &  21.32  \\ %negative LBZ?? %HARM3D/MAD/harm_mad482/exe/Model6a_3D
      GRB-LS100  & 6562  & 2443 & 0.3 & 6.0 & 13.0 & 100 & 0.0319 & 0.082 & 2.191 & 0.0004 \\ %/net/scratch/people/plgagnesj/harm/harm_mad482/exe/Model3b2_a03
      GRB-LS50 & 6562  & 2392 & 0.3 & 6.0 & 13.0 & 50 &   0.0228 & 0.075 & 2.359 & 0.0007 \\ %/net/scratch/people/plgagnesj/harm/harm_mad482/exe/Model3b1_a03
      GRB-LS30 & 6562  & 2534 & 0.3 & 6.0 & 13.0 & 30 &   0.0105 & 0.065 & 2.284 & 0.0013 \\ %/net/scratch/people/plgagnesj/harm/harm_mad482/exe/Model3b1_a03
      \hline\hline
       AGN-HS100 & 532647  & 309765 & 0.9 & 12.0 & 25.0 & 100 & 0.120 & 3.765 & 8.801 & $8.3\times10^{-5}$\\ %/net/scratch/people/plgagnesj/harm/harm_mad482/exe/Model5_rmax25_a09_beta100%
      AGN-HS50 & 532647   & 274132 & 0.9 & 12.0 & 25.0 & 50 & 0.109 & 3.465  &  9.177 & $2.5\times 10^{-3}$ \\ %HARM3D/MAD/harm_mad482/exe/Model5a_rmax25_3D%
         AGN-HS30 & 532647   & 293182 & 0.9 & 12.0 & 25.0 & 30 & 0.069 & 3.104 &  8.129 &  $2.9\times 10^{-3}$ \\ %HARM3D/MAD/harm_mad482/exe/Model6a_rmax25_3D%
      AGN-NS100 & 298344  & 195455 & 0.0 & 12.0 & 25.0 & 100 & -- & 2.284 & 8.421 & $4.9\times 10^{-5}$  \\%/net/scratch/people/plgagnesj/harm/harm_mad482/exe/Model4_rmax25_beta100/
      AGN-NS50 & 298344  & 210646 & 0.0 & 12.0 & 25.0 & 50 &  -- & 1.976 & 6.442 & $6.1\times 10^{-5}$ \\ %/net/scratch/people/plgagnesj/harm/harm_mad482/exe/Model4_rmax25_beta50/
      AGN-NS30 & 298344  & 210083 & 0.0 & 12.0 & 25.0 & 30 & -- & 2.043 & 6.197 & $7.2\times10^{-5}$  \\ %/net/scratch/people/plgagnesj/harm/harm_mad482/exe/Model4_rmax25_beta30/
%      Magn1 & 0.1104  & 0.0895 & 0.9 & 4.5 & 9.1 & 1.e-2 & $8.76\cdot 10^{-2}$ & 0.00066\\
      %      Magn2 & 0.1104   & 0.0682 & 09  & 4.5 & 9.1 & 1.e-1 & $1.20\cdot 10^{-1}$ &  0.00076
%      \enddata
      \hline
      \end{tabular}
\caption{ Summary of the models. 
  The inner radius of the torus, $r_{\rm in}$, the radius of pressure maximum, $r_{\rm max}$, and the plasma-$\beta$ are given as the initial state parameters. The last two columns give the magnetic flux %normalized to accretion rate at inner boundary,
  averaged over
  the simulation time, and the total 
  mass lost through the outer boundary.
  All quantities are given in code units. The BZ power is given at the mid-time of simulation,
  at~ $[25000 t_{g}]$.
  The final time for all simulations is 50000 $t_{g}$ }
\label{tab:in}
   \end{table*}

\subsection{General structure of the accretion flow}

Our simulations are divided into two sets with the \textit{GRB} class of models referring to a system composed of a stellar mass black hole surrounded by a remnant accretion disk formed following a compact binary merger, while the \textit{AGN} class of models refer to an engine of active galaxy, where a supermassive black hole accretes matter of a diluted torus in the circum-nuclear region. The second class of models adopts larger spatial extension of the tori, in terms of their pressure maximum radius. The inner radius of the torus is also located further out from the black hole horizon in AGN models (in terms of the dimensionless units). We further consider fast and slow spinning black holes, including non-spinning ones for an AGN case to describe radio-quiet sources. Finally, we also vary the initial plasma $\beta$-parameter.

Table~\ref{tab:in} presents a summary of the models and their initial parameters, the total mass of the disk at initial state, the averaged mass accretion rate, the average magnetic flux of the black hole horizon, and the total mass lost in the outflows. We also calculate the jet power at the mid-state of the simulations.

The overall evolution pattern is as follows. The FM torus initial state retains its structure for a relatively long integration time. Its innermost, densest part
is enclosed with $\sim 50 r_{g}$ and forms a narrow cusp through which the matter sinks under the black hole horizon. The surface layers of the torus, which extend to about 200 $r_{g}$, form a kind of corona. This region is the base of
sub-relativistic outflows that are launched from the center as soon as the initial condition of the simulation is relaxed (which is at about 2000 M).
The models were evolved until the time of $t_{\rm f}=50,000$ $t_{g}$ (geometrical units, $t_{g}=G M_{\rm BH}/c^{3}$). 
This corresponds to 0.74 seconds in the GRB case and to about 9.4 months in our AGN case.

\begin{figure*}
\begin{tabular}{ccc}
   \hspace{-10mm}\includegraphics[width=0.39\textwidth]{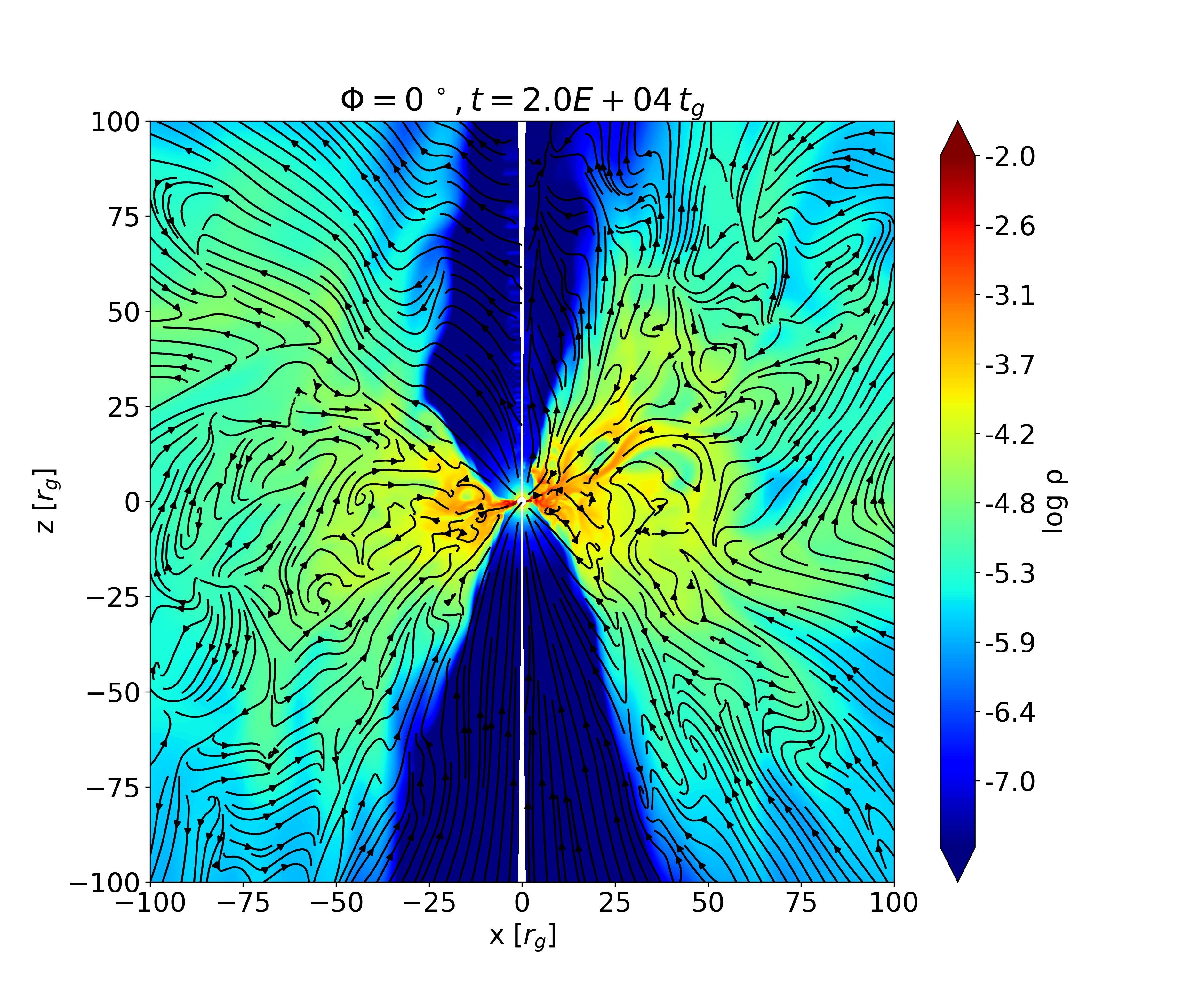} &
  \hspace{-10mm}\includegraphics[width=0.39\textwidth]{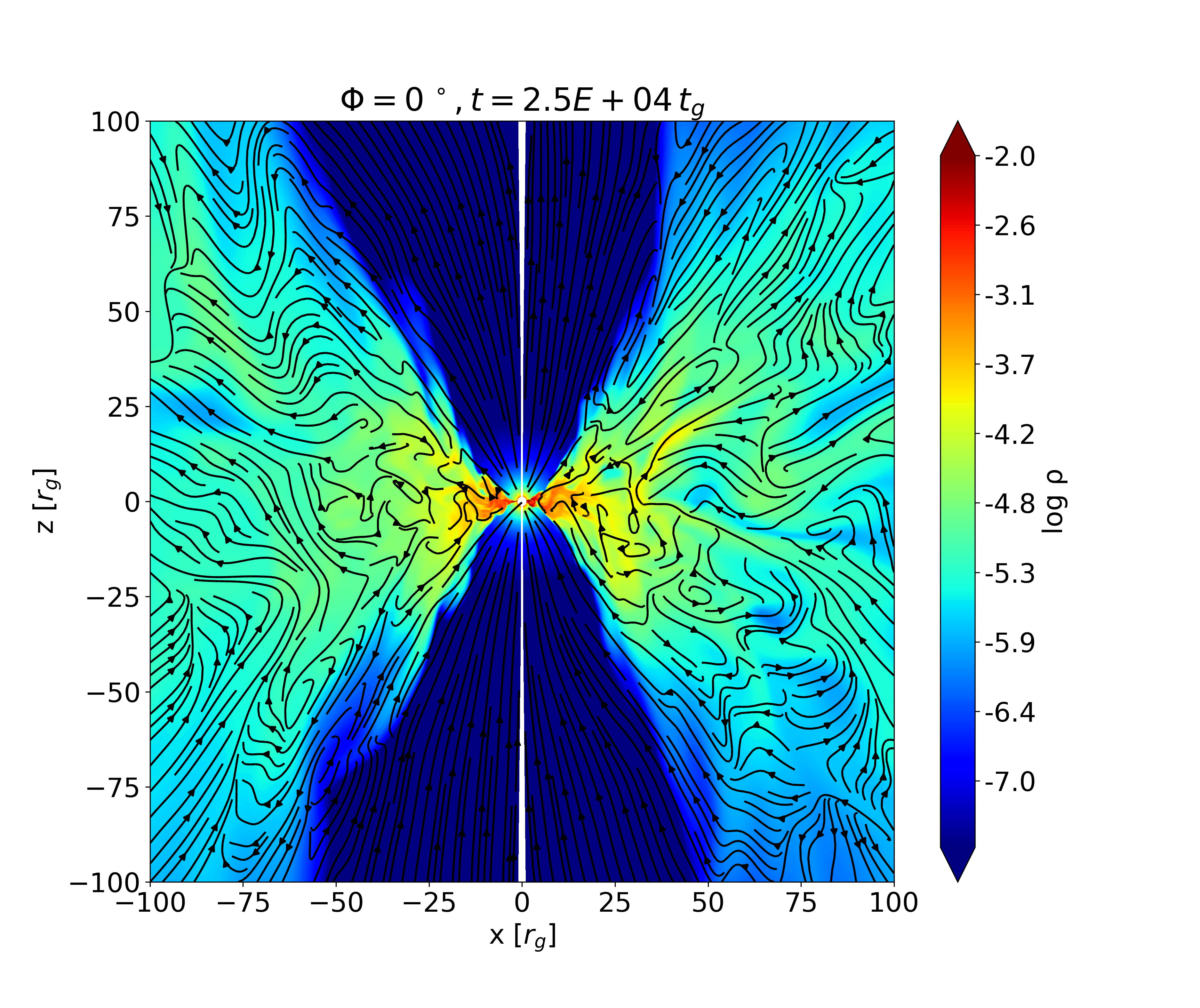} &
  \hspace{-10mm}\includegraphics[width=0.39\textwidth]{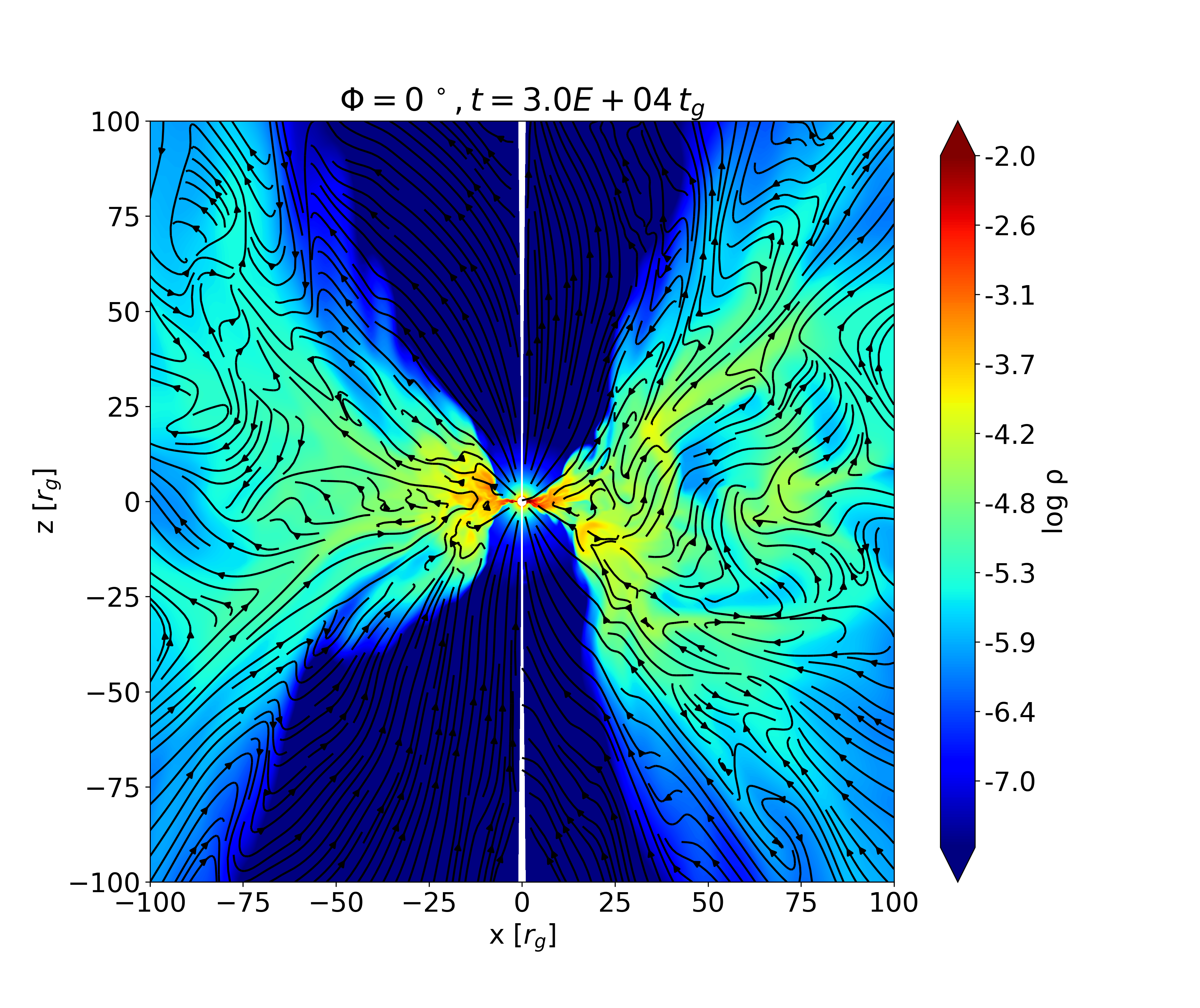}
\\  
   \hspace{-10mm}\includegraphics[width=0.39\textwidth]{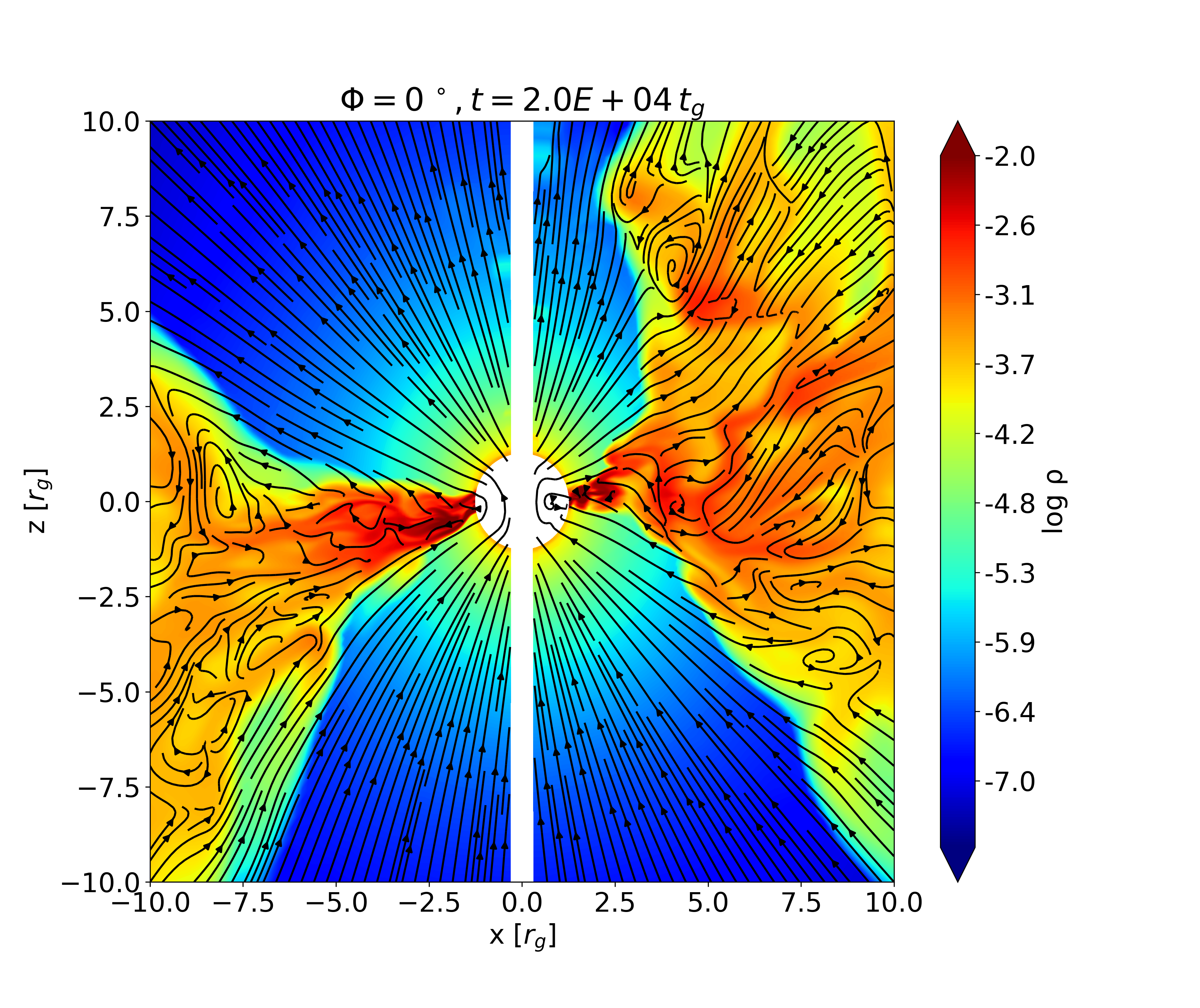} &
  \hspace{-10mm}\includegraphics[width=0.39\textwidth]{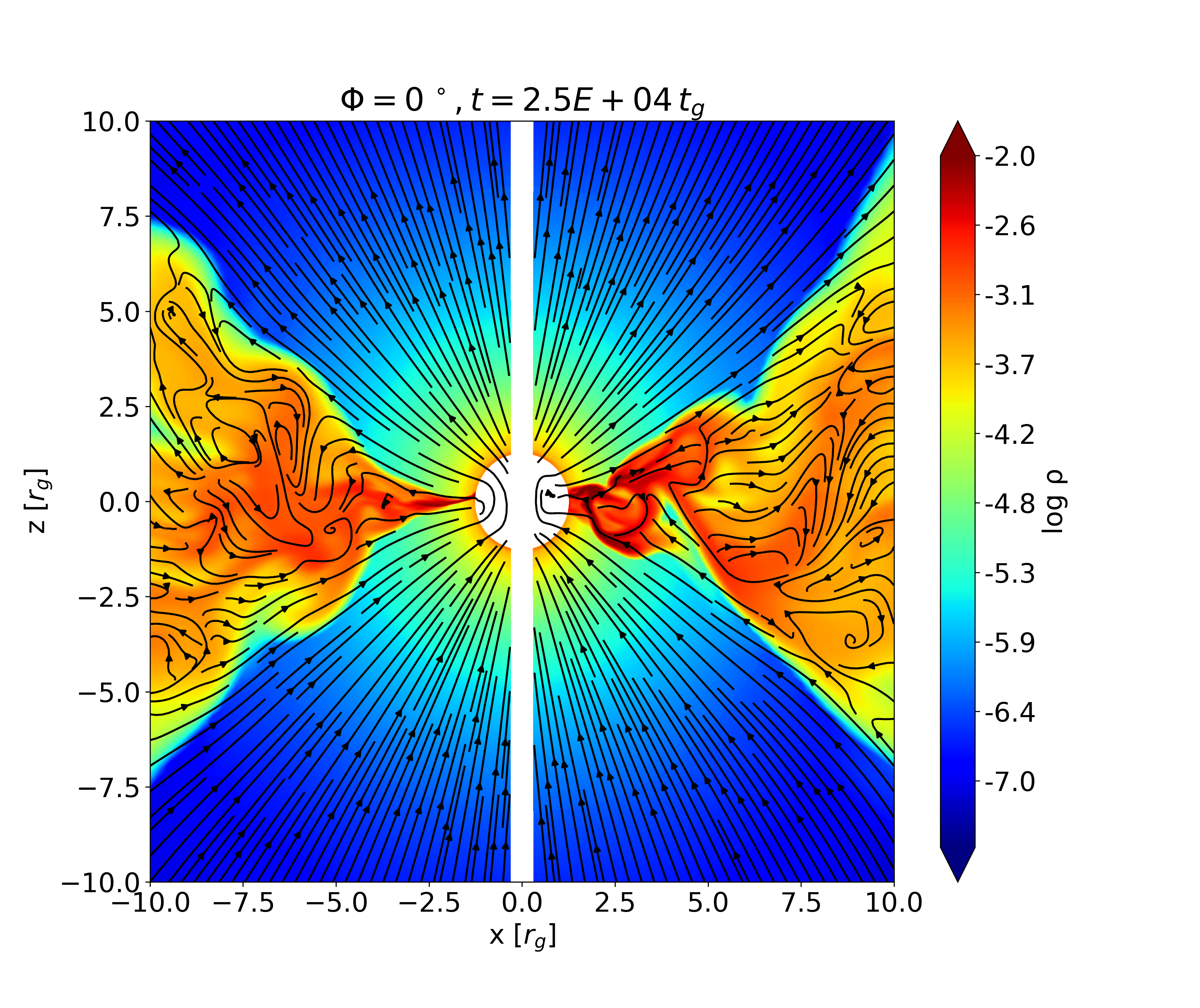} &
  \hspace{-10mm}\includegraphics[width=0.39\textwidth]{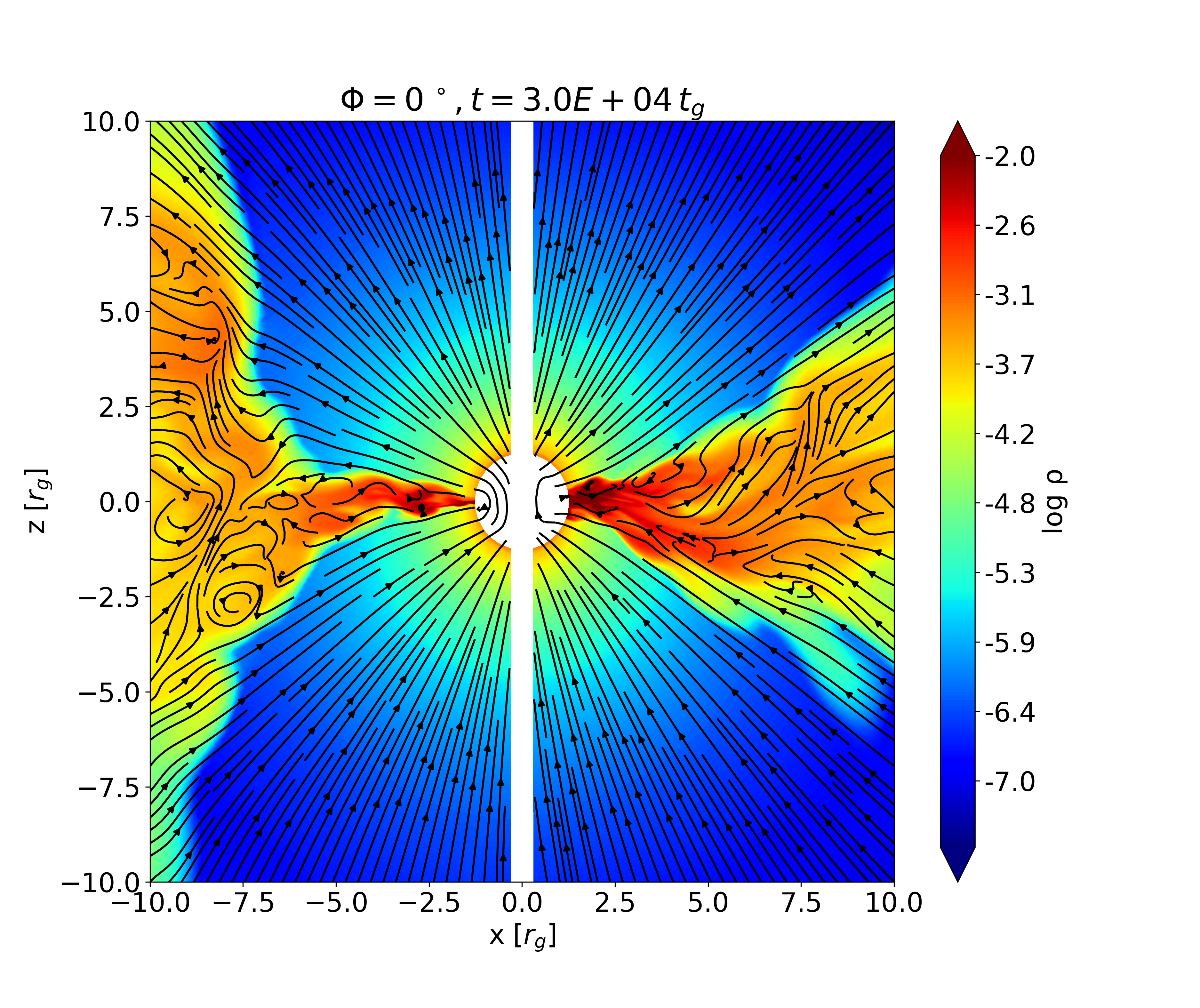} 
\\
  \hspace{-12mm}\includegraphics[width=0.39\textwidth]{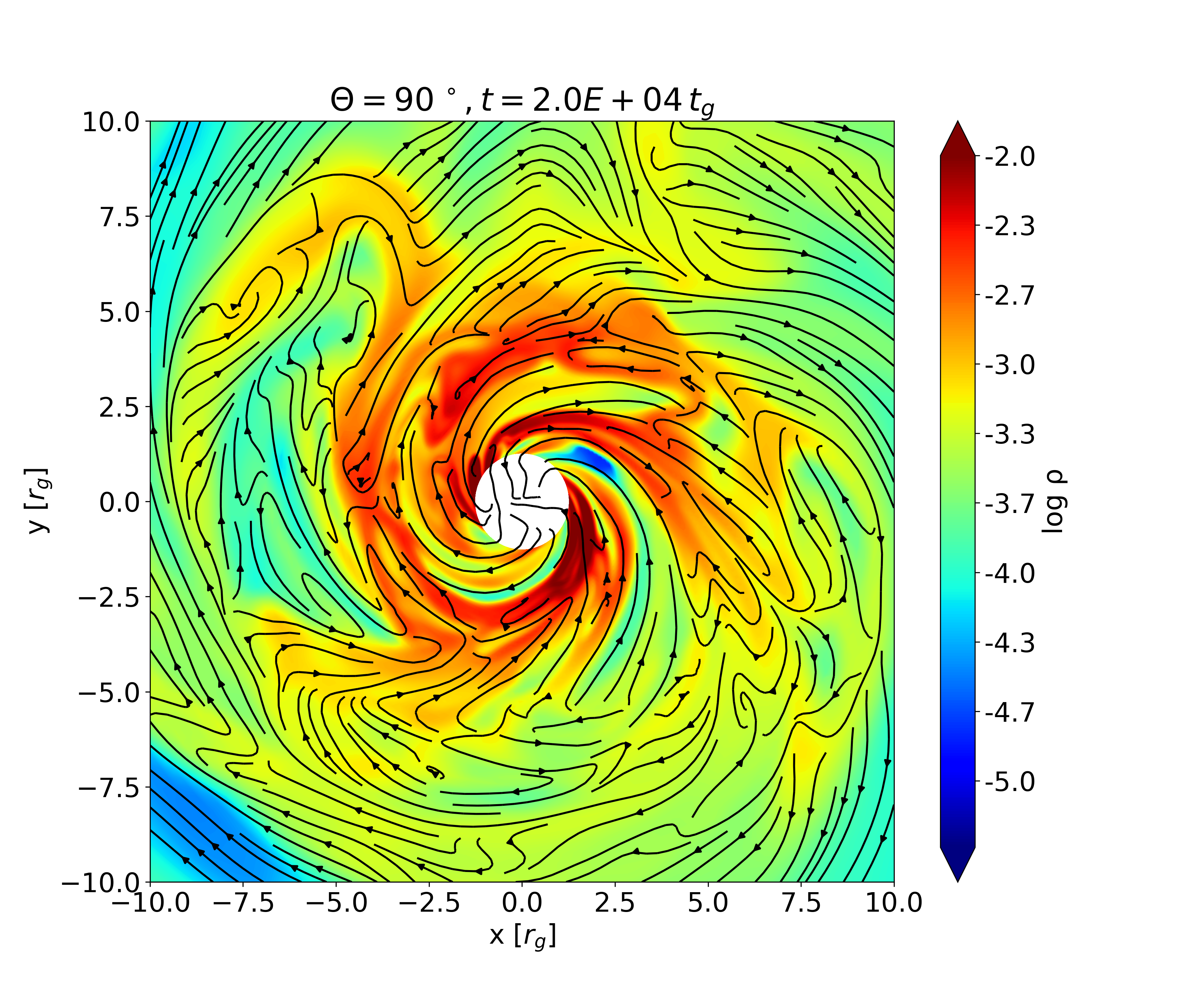}  &
  \hspace{-12mm}\includegraphics[width=0.39\textwidth]{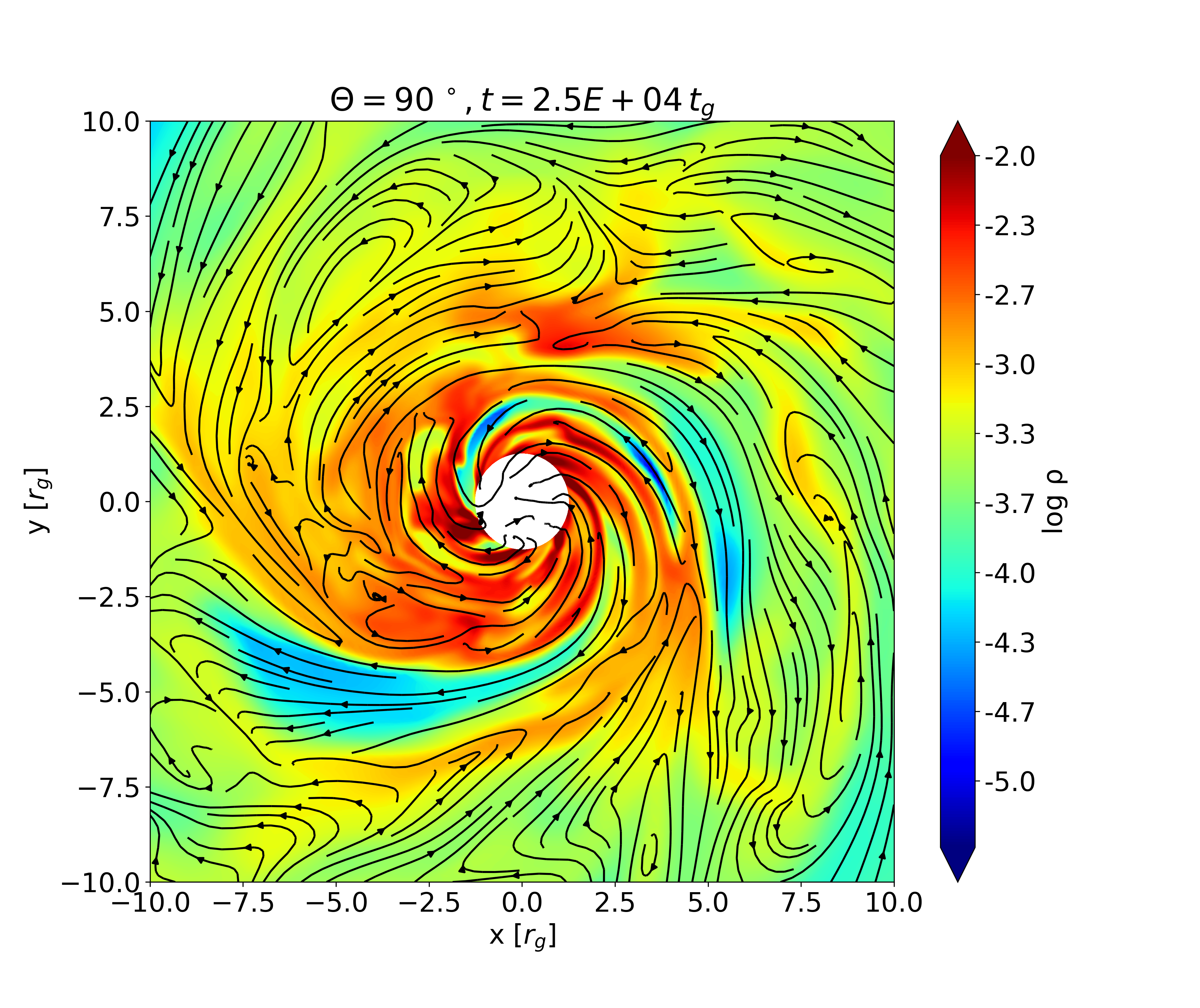}  &
    \hspace{-12mm}\includegraphics[width=0.39\textwidth]{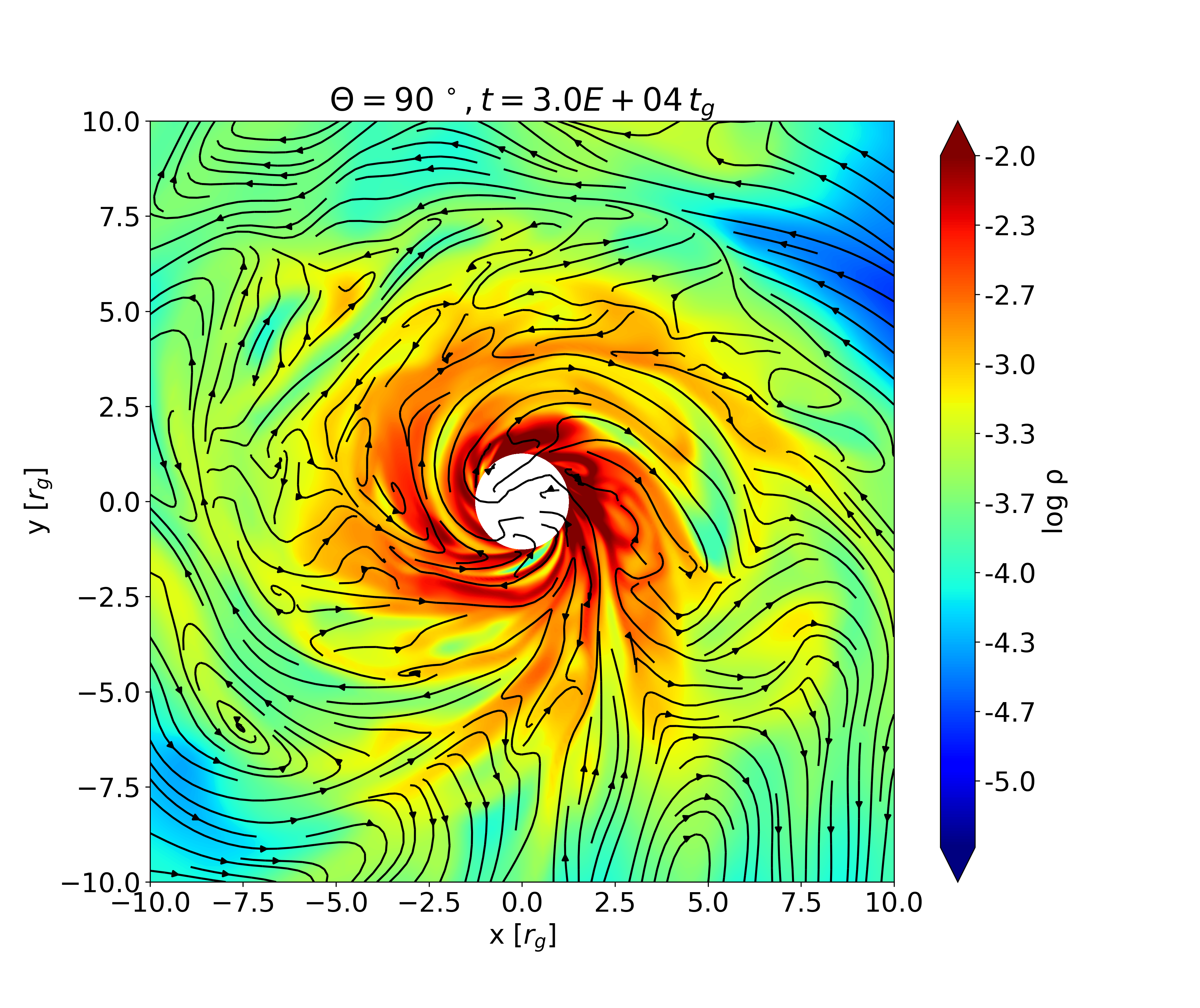} 
\end{tabular}
\caption{Maps of the density isocontours in the equatorial plane XY (bottom row), and polar plane XZ (top row). The model is $GRB-HS100$ and its parameters are $\beta=100$ and $a=0.9$.
  The color maps are in logarithmic scale and are taken at the three times
  of the simulation, at  $t=20,000 M$, $t=25,000 M$, and  $t=30,000 M$.
  %(i.e., $\sim 0.3$ second for $M_{\rm BH}=3 M_{\odot}$). 
  Streamlines show the magnetic three-vector configuration.
   We show the equatorial cuts of a small region of $10 r_{g}$ from the black hole (bottom panels) and the polar cuts
  are shown for small and large regions, up to $100 r_{g}$ (upper panels). 
 }
    \label{fig:models_results1_HS}
\end{figure*}

In Figure \ref{fig:models_results1_HS}, we show the profiles of density, and magnetic field streamlines,
in the $r-\theta$ polar plane (XZ), and in the equatorial $r-\phi$ plane (XY),
as obtained at the mid-time of the simulation.
 We show three different times during the simulation, $t=20,000 M$ $t=25,000 M$, and $t=30,000 M$, in the left, middle, and right panels, respectively. This sequence of snapshots shows that the dense torus wobbles about the equatorial plane, while the jet funnel is slightly asymmetric with respect to the vertical
  axis.
The model shown here is \textit{GRB-HS100}, representing a weakly magnetized, small torus with $\beta=100$ and black hole spin of $a=0.9$.
The  top panels are scaled to distance of 100 $r_{g}$, while the middle and bottom panels show a zoom-in on the accretion region, within the inner 10 gravitational radii. 
As shown in these maps, the magnetically arrested flow appears
in this simulation, forcing the accretion to proceed via the azimuthally distributed "fingers" surrounding the  black hole and entering its horizon. The zoomed-in equatorial plots show that
the  magnetic field lines are pulled into black hole within the low-density regions, while the high-density fingers are pushing out the magnetic fields. Thus, the magnetic tension can locally push the gas out and suppress the accretion rate, while in denser regions, the matter prevents the flux from entering the horizon. We checked that the total mass accretion rate integrated over the horizon region in these three time intervals is equal to 0.0208, 0.0224, and 0.0110, respectively, while the magnetic flux integrated over the horizon is 1.11, 1.53, and 1.35 (in code units). Hence, the smallest net accretion rate and highest rate characterizing the magnetically arrested state are found in the final time snapshot shown.
It is only very high resolution simulations (see e.g., \citep{Ripperdaetal}) that can verify whether the MAD regions with the dense magnetic flux getting pushed out can serve as sites of magnetic reconnection. Our simulations have  moderate resolution in 3D, so that we encompass this effect only qualitatively.
We also note that when the jet forms, hot luminous regions appear near the black hole rotation axis, which have a very low baryon density and a vertically oriented magnetic field. These polar regions represent bipolar jets, with a luminosity that depends on the black hole spin (cf. Table \ref{tab:in}).

 In the model of an AGN disk around a non-spinning black hole,
the weakly magnetized torus is of a larger size and its inner radius and radius of pressure maximum are 12 and 25 $r_{g}$, respectively. We may notice that in this model, the large mass of the torus  and zero black hole spin
does not allow for the formation of magnetically dominated, low-density funnels at the polar region.
 The non-spinning black holes in AGN engines do, however, also accrete in the MAD mode. As shown recently by \cite{ipole2021}, the specific features of this accretion mode lead to the polarized radiation emission from the region of the black hole horizon. Here, in contrast to the emission from remote jets, the magnitude of the circular polarization is high. Because of high Faraday thickness, the circular polarization is particularly sensitive to dynamics of toroidal magnetic field in the accretion flow.

The case of  a fast-rotating black hole engine of a gamma ray burst, with smaller values of initial $\beta$, equal to 50, and 30, was studied in 
models \textit{GRB-HS30} and \textit{GRB-HS50}. Here, also the polar funnel and equatorial
  disk structure have been developed, with accretion proceeding through the equator, while bi polar, magnetically dominated jets are launched.
  The qualitative differences between these runs and $\beta=100$ case are seen in the density range in the equatorial plane, which is (on average) larger than small $\beta$ models. The clear asymmetry in the azimuthal distribution of matter, which is otherwise developing in the \textit{GRB-HS30} model, results from stronger magnetic fields which are pushing the accretion flow out from the horizon. The barrier shape is also not symmetric, so the gas must overpass it and reach the black hole on the opposite side.
 
 \subsection{Mass inflow and the MAD\ state}

 \begin{figure}
    \centering
     \includegraphics[width=0.45\textwidth]{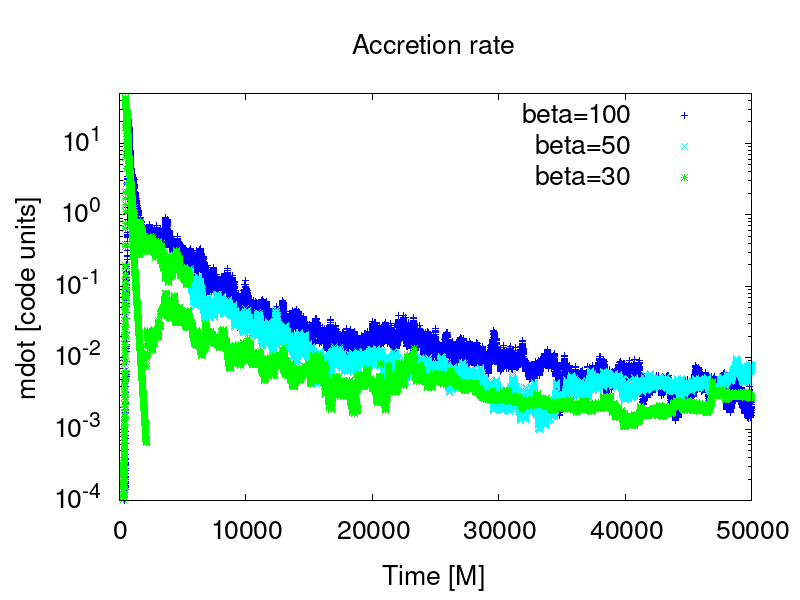}
     \includegraphics[width=0.45\textwidth]{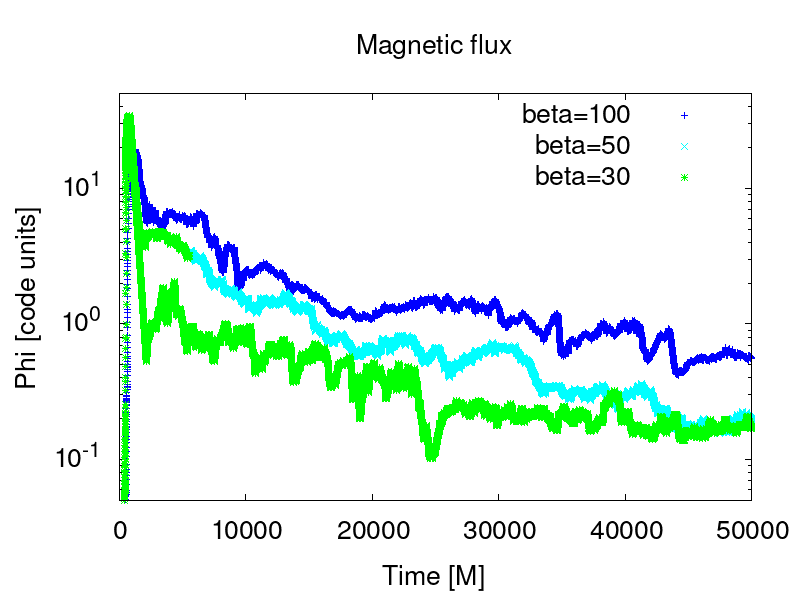}
      \includegraphics[width=0.45\textwidth]{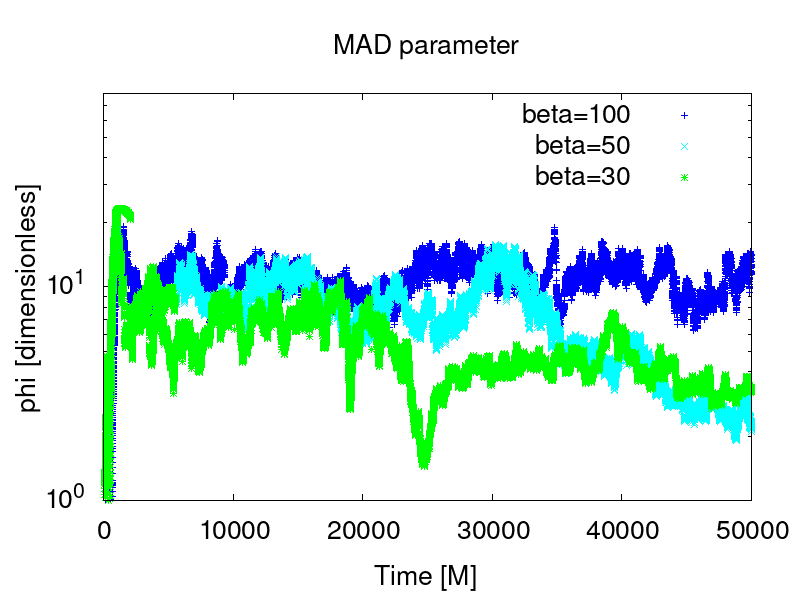}
      \caption{Time dependence of the mass flux through the black hole horizon (top panel), horizon-integrated magnetic flux (middle panel), and the so-called MAD-ness parameter, $\phi_{BH}$, given by Eq. \ref{eq:madness} (bottom panel).
        Blue lines denote the models with initial
        gas-to-magnetic pressure ratio, $\beta=100$, cyan lines present the results for $\beta=50$,
        and green lines denote $\beta=30$. Black hole spin parameter in all models is $a=0.9$. The disk inner radius is located at 6 and maximum pressure radius at 13 $r_{g}$ (GRB-type models).
      }
     \label{fig:timedep1a2a}
     \end{figure}

 In Figure \ref{fig:timedep1a2a}, we show the time-dependence of so called MAD-ness parameter, namely, the ratio of the mass and magnetic flux through the black hole horizon.
 These are given by the relations:
 \begin{equation}
{\dot m}_{BH} = - \int{\rho u^{r} \sqrt{-g} d\theta d\phi}
 ,\end{equation}
 \begin{equation}
{\Phi}_{BH} = 0.5 \int{|{B^{r}}| \sqrt{-g} d\theta d\phi}
 ,\end{equation}
 and
  \begin{equation}
    {\phi}_{BH} = {\Phi_{BH} \over \sqrt{\dot{m}_{BH}}}
    \label{eq:madness}
 .\end{equation}
  As shown in Figure \ref{fig:timedep1a2a}, 
  accretion rate is large at the beginning of the simulation and it is higher then for smaller $\beta$ value.
  At later times, the magnetic flux that has been dragged to the black hole vicinity, growing high enough to capture the accretion flow. Hence, the ratio of these two fluxes remains constant and greater than $\phi_{BH}\sim 10$ for most of the simulation after the initial condition is relaxed. The quantitative difference in $\phi_{BH}(time)$ profile between the models with $\beta=100$ and $\beta=50$ is very small in the case of the average value.
  However, the time variability of the more magnetized flow appears to be increased; this relation is quantified in more detail below (i.e., Sect. \ref{sec:variability}).

\begin{figure}
    \centering
     \includegraphics[width=0.45\textwidth]{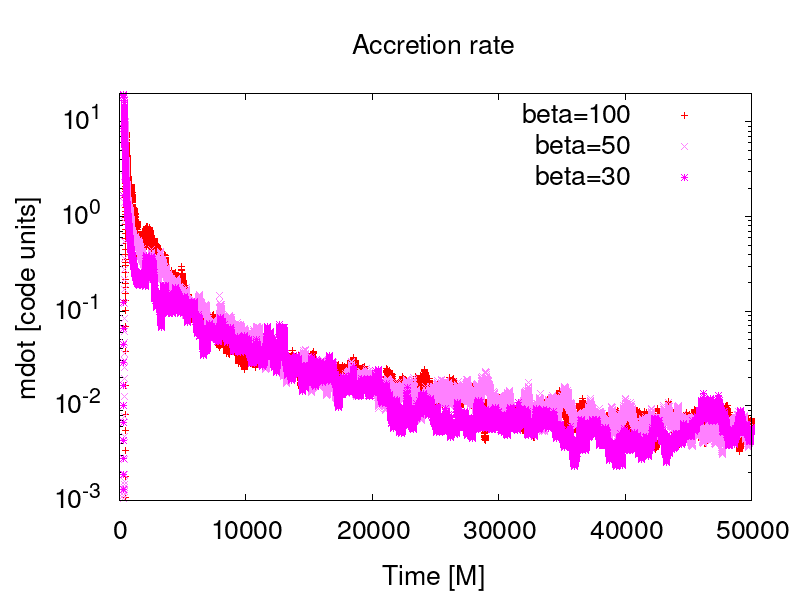}
     \includegraphics[width=0.45\textwidth]{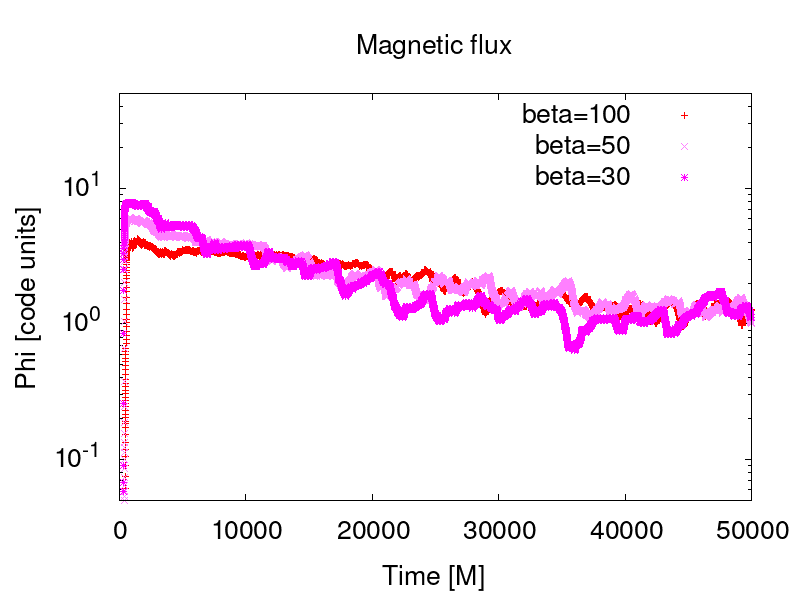}
      \includegraphics[width=0.45\textwidth]{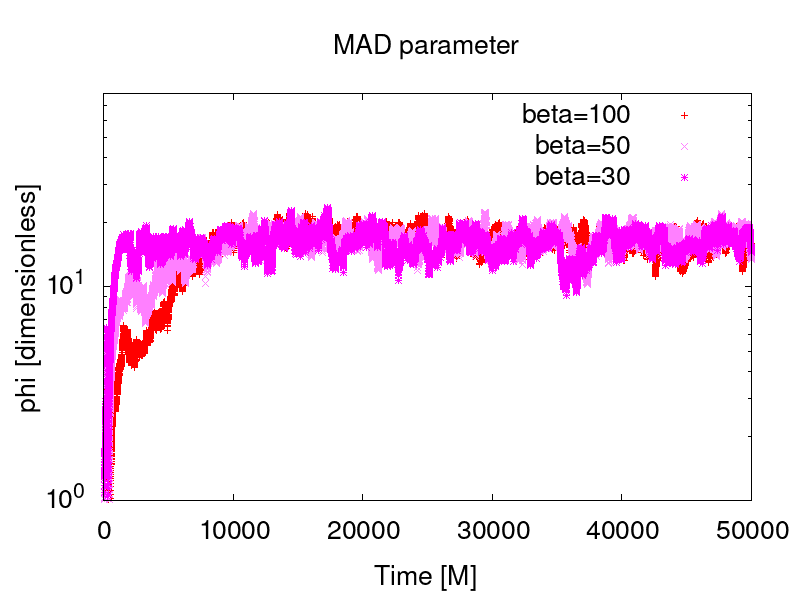}
      \caption{Time dependence of the mass flux through the black hole horizon (top panel), the horizon integrated magnetic flux (middle panel), and the so-called MAD-ness parameter, $\phi_{BH}$, given by Eq. \ref{eq:madness} (bottom panel).
         Red lines denote the models with initial
        gas-to-magnetic pressure ratio $\beta=100$,  orchid lines present results for $\beta=50$,
        and  magenta lines denote $\beta=30$. Black hole spin parameter in all models is $a=0.3$. The disk inner radius is located at 6 and maximum pressure radius at 13 $r_{g}$ (GRB-type models).
      }
     \label{fig:timedep03}
     \end{figure}
     
In Figure \ref{fig:timedep03}, we present the results of simulation with a black hole spin of $a=0.3$ and the same small accretion disk, with $r_{in}=6$ and $r_{max}=13$ $r_{g}$, referring to the \textit{GRB-LS100}, \textit{GRB-LS50}, and \textit{GRB-LS30} models. The three values for the $\beta$-parameter were chosen to differentiate between the  magnetic field strength in these models, which are plotted with red, orchid, and magenta lines.
We observe that for this small value of black hole spin, the magnetic flux on the horizon, $\Phi_{B}$, is almost the same for all three values of $\beta$. It is also more stable than it was in the high-spin models, so that it has a smaller value at the beginning of the run, but does not drop so much over time, as compared to the models with a spin of $a=0.9$. 
As the  time evolution for the accretion rates is rather similar in both these simulation runs, the average profile of MAD-ness parameter remains almost constant with time, namely, at the level of $\phi_{BH}=20$.
The timescales of variability seem to be rather similar to each other and not depend much on the magnetization parameter (see, however, the quantitative analysis in Sect. \ref{sec:variability}.)

\begin{figure}
    \centering
     \includegraphics[width=0.45\textwidth]{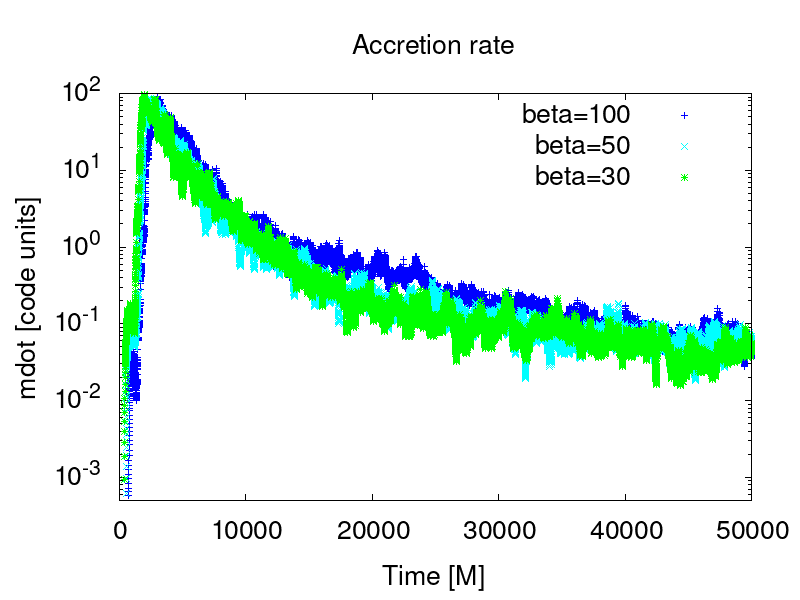}
     \includegraphics[width=0.45\textwidth]{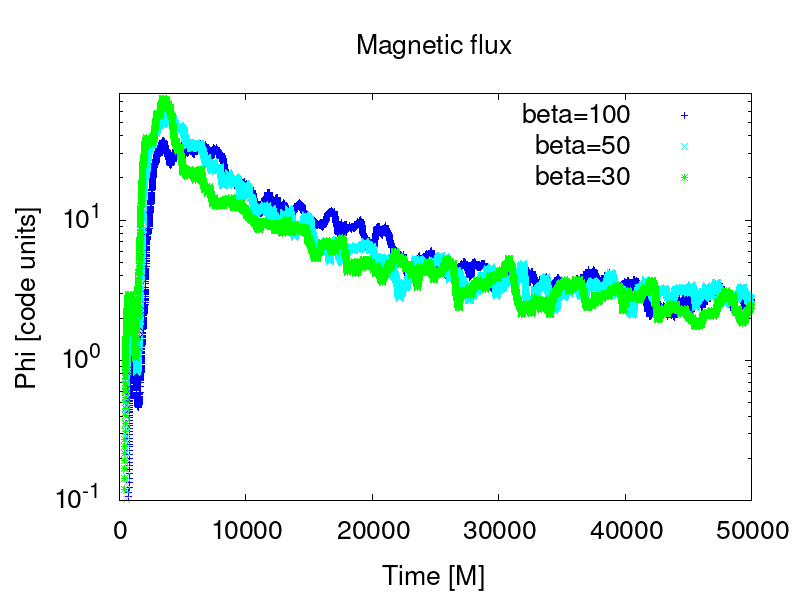}
      \includegraphics[width=0.45\textwidth]{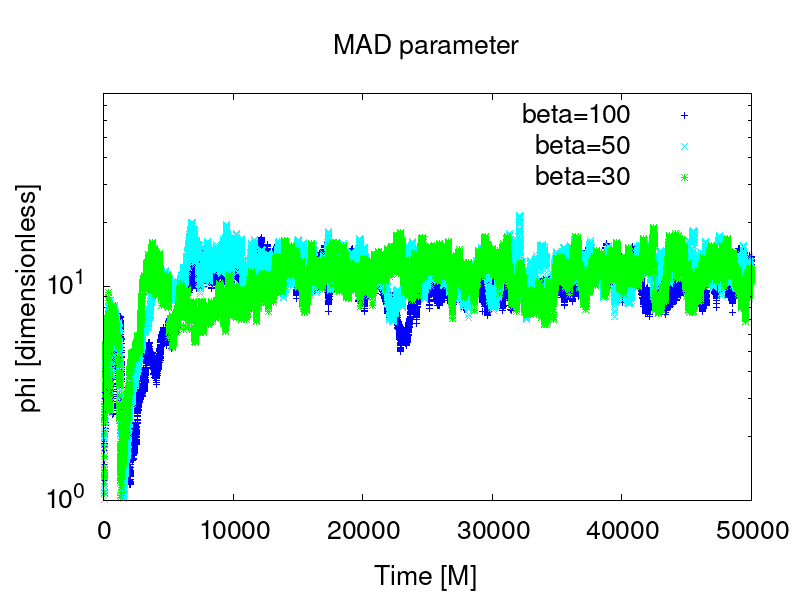}
      \caption{Time dependence of the mass flux through the black hole horizon (top panel), the horizon integrated magnetic flux (middle panel), and the so-called MAD-ness parameter, $\phi_{BH}$, given by Eq. \ref{eq:madness} (bottom panel).
        Blue lines denote the models with initial
        gas-to-magnetic pressure ratio $\beta=100$, cyan lines present results for $\beta=50$,
        and green lines denote $\beta=30$. Black hole spin parameter in all models is $a=0.9$. The disk inner radius is located at 12 and maximum pressure radius at 25 $r_{g}$ (AGN-type models).
      }
     \label{fig:timedep09d}
     \end{figure}
     
     Finally, we present in Figure \ref{fig:timedep09d}  a similar set of models, assuming, for the black hole spin, $a=0.9,$ and three-disk magnetization, normalized to $\beta=100$, 50, and 30, while embedding the larger torus. The radii of inner disk cusp and its pressure maximum are taken as $r_{in}=12$ and $r_{max}=25 r_{g}$, respectively. 
Now, both the mass accretion rate and magnetic flux on the horizon, remain large for the time of the simulation. The MAD state is prolonged, while the MAD-ness parameter, $\phi_{BH}$, rises smoothly after an initial relaxation period of about a 1000~t$_{g}$. The differences between the results obtained for three values of $\beta$-parameter are now less pronounced. 
From the above three sets of simulations, we conclude therefore that the effect of magnetization changing the variability timescale depends on the black hole spin and the size and position of the torus. In our GRB models with high values for the black hole spin, the variability pattern is affected by the disk magnetization. In models with either small spin or a large torus, referring to the AGN case, the magnetic fields do not seem to alter the variability timescales as significantly.
   
       \subsection{Variability analysis}
       \label{sec:variability}

     We  quantitatively analyzed the variability of the accretion rate onto the black hole by means of a Fourier transformation.

\begin{figure}
    \centering
     \includegraphics[width=0.45\textwidth]{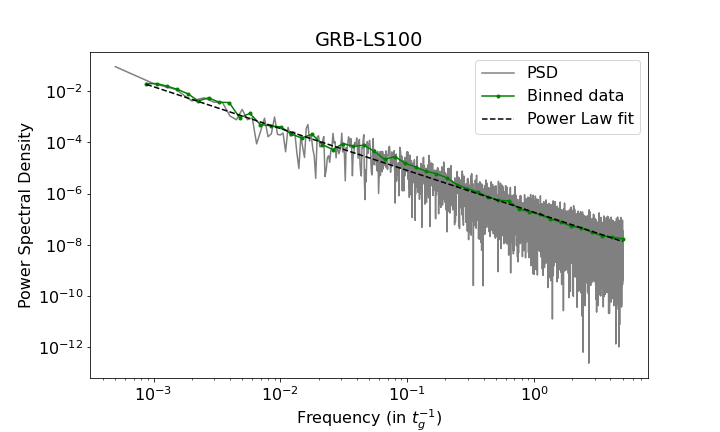}
     \includegraphics[width=0.45\textwidth]{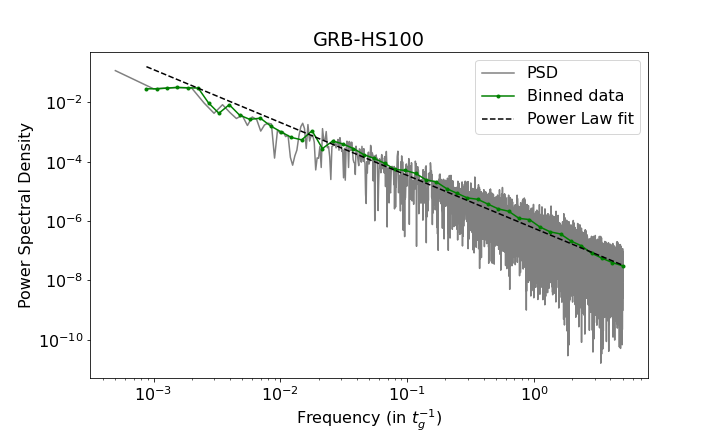}
     \includegraphics[width=0.45\textwidth]{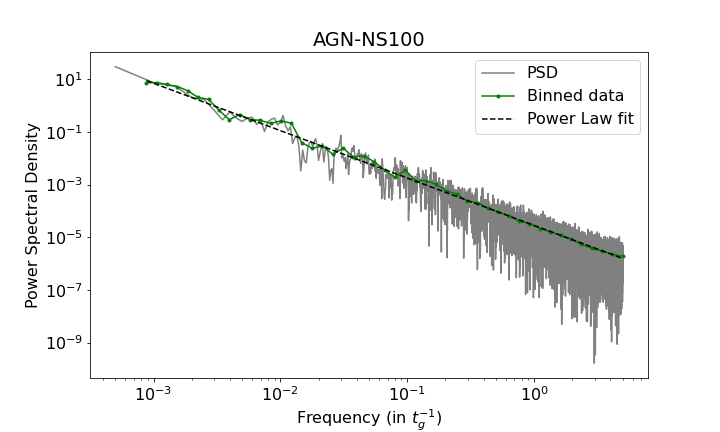}
     \includegraphics[width=0.45\textwidth]{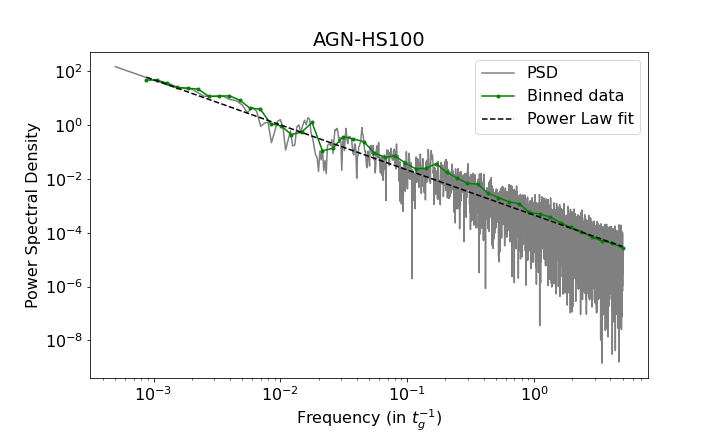}
      \caption{
        PDS and their power-law fits presented for four exemplary simulations. The models are: GRB-LS100, GRB-HS100, AGN-NS100, and AGN-HS100, as marked in the figures. The light curves were binned logarithmically to obtain better fits.
      }
      \label{fig:pds}
     \end{figure}

     We calculated the power-density spectra of the time-series of \textit{GRB} models, depicted in Figures \ref{fig:timedep1a2a} and  \ref{fig:timedep03}; and for \textit{AGN} models, shown in Fig. \ref{fig:timedep09d}, as well as for non-spinning black hole models.
     In the analysis, we first skipped $\delta t=10,000$~t$_{g}$ and we focused on the variability of the flow when the initial transient phase is relaxed. Figure \ref{fig:pds} shows the power density spectra of four chosen models, with the smallest initial magnetization in the disk.
     
     Table \ref{tab:pds_results} gives the values of power-law fitting to the power density spectra (PDS) of individual models. We notice that the PDS slope depends on the black hole spin, while only weakly on the disk magnetization. For both classes of models, GRB and AGN, the PDS slopes tend to be steeper for lower spin of the black hole.
     For highly spinning black hole models, the GRB models present flatter PDS spectra than the AGN models.
     We also calculated the minimum variability timescale, estimated as the minimum peak width at its half maximum, over the time series (here, we also skipped the first $10,000$~t$_{g}$). There is an anticorrelation between this timescale and the disk magnetization for GRB and AGN models. Also, models with highly spinning black holes tend to present shorter variability timescales than those with slowly or non-spinning black holes.
   
     %5-7.
          
\begin{table}
\centering
\begin{tabular}{lcccr}
  %\toprule
  \hline
  \hline
Model & a & Initial $\beta$ & PDS slope & MTS \\
\hline
\hline
 GRB-HS100 & 0.9 & 100 & -1.78 $\pm 0.02$ & 29.15 \\
 GRB-HS50  & 0.9 & 50  & -1.67 $\pm 0.02$ & 30.95 \\
 GRB-HS30  & 0.9 & 30  & -1.97 $\pm 0.01$ & 39.18 \\
 \hline
 GRB-LS100 & 0.3 & 100 & -1.64 $\pm 0.03$ & 70.06 \\
 GRB-LS50  & 0.3 & 50  & -1.80 $\pm 0.01$ & 66.64 \\
 GRB-LS30  & 0.3 & 30  & -1.66 $\pm 0.03$ & 62.79 \\
 \hline
 AGN-HS100 & 0.9 & 100 & -1.68 $\pm 0.01$ & 29.95 \\
 AGN-HS50  & 0.9 & 50  & -1.68 $\pm 0.01$ & 30.78 \\
 AGN-HS30  & 0.9 & 50  & -1.70 $\pm 0.01$ & 26.64 \\
 \hline
 AGN-NS100 & 0.0 & 100 & -1.80 $\pm 0.02$ & 72.84 \\
 AGN-NS50  & 0.0 & 50  & -1.66 $\pm 0.03$ & 90.63 \\
 AGN-NS30  & 0.0 & 30  & -1.87 $\pm 0.02$ & 77.39 \\
 \hline
 
\end{tabular}
\caption{Results of power density spectral fitting. Model parameters are black hole spin $a$ and $\beta=p_{gas}/p_{mag}$. The PDS slope is fitted power law index. MTS is the minimum variability timescale, in $t_{g}$ units.}
\label{tab:pds_results}
\end{table}
     
     \subsection{Jet power and energetic structure}

     The jets in GRB models are Poynting energy-dominated and confined by the toroidal magnetic fields. Figure \ref{fig:map_Bspiral} shows the 3D structure of the magnetic field, as depicted by an arbitrarily chosen field line, tangential to the magnetic field vectors and footed at the black hole horizon. The model shown here is \textit{GRB-HS100} with black hole spin $a=0.9$.
 
In Table \ref{tab:in}, we give the values of jet power, as calculated in code units in the mid-time of the simulation,  at $25,000 ~ t_{g}$. 
The power is computed  using an estimate of the Blandford-Znajek process, where the rotating black hole is surrounded by a force-free, magnetized plasma. We use the formula given by \cite{McKinney2004}:
\begin{equation}
P_{\rm BZ}= - \int_{r=r_{h}} (b^{2}u^{r}u_{t} - b^{r}b_{t}) dA
\label{eq:power}
,\end{equation}
so that we integrate the radial part of electromagnetic flux over the surface of black hole horizon.
In general, the integral over the radial energy flux, $T^{r}_{t}$, may be subdivided into matter and electromagnetic parts. In the above formula, we neglect the matter component, assuming a force-free approximation. For a steady, uniform outflow, this expression gives  the radial Poynting energy flux measured by a stationary observer at large distance from the black hole. We notice that the BZ power is positive only if the energy can be extracted to the remote jets from the black hole, so only the flux with a positive sign
will contribute to it. Otherwise, there is no energy extraction from the black hole via the BZ process.
These jets, if produced, will be dominated by thermal energy.

 As listed in Table \ref{tab:in}, at the mid-time of the simulation, a strong, Poynting-dominated bipolar jets have already developed in most of the models. We note that for the models with zero black hole spin, such jets have not emerged; therefore, the value computed from  Eq. \ref{eq:power} is set to zero.
We also notice that in the model \textit{GRB-HS30}, with a high black hole spin,
 the Poynting-dominated jet is quenched.
The polar funnels are having a low density, however, the MAD state is not operating, according to a standard definition, with $\phi_{BH}$ value being smaller than 10.

For all other simulations, the jet power obtained in the Blandford-Znajek process  is non-zero for the black hole spin, $a>0,$ and  at the mid-term of the simulation, it ranges from $\sim 10^{42}$ to $\sim 10^{52}$ erg s$^{-1}$, for AGN and GRB, respectively. We calculate this power 
and we normalize it with the adopted physical scalings, related to the black hole mass.
The characteristic quantities in physical units and unit conversions, are summarized in Table \ref{tab:conversion}. We note that the jet power in GRB scalings is very large and it could be relevant only for some extremely bright sources. It is thus possible that the MAD scenario is not capable of explaining faint GRBs.

\begin{figure}
\centering
    \includegraphics[width=0.45\textwidth]{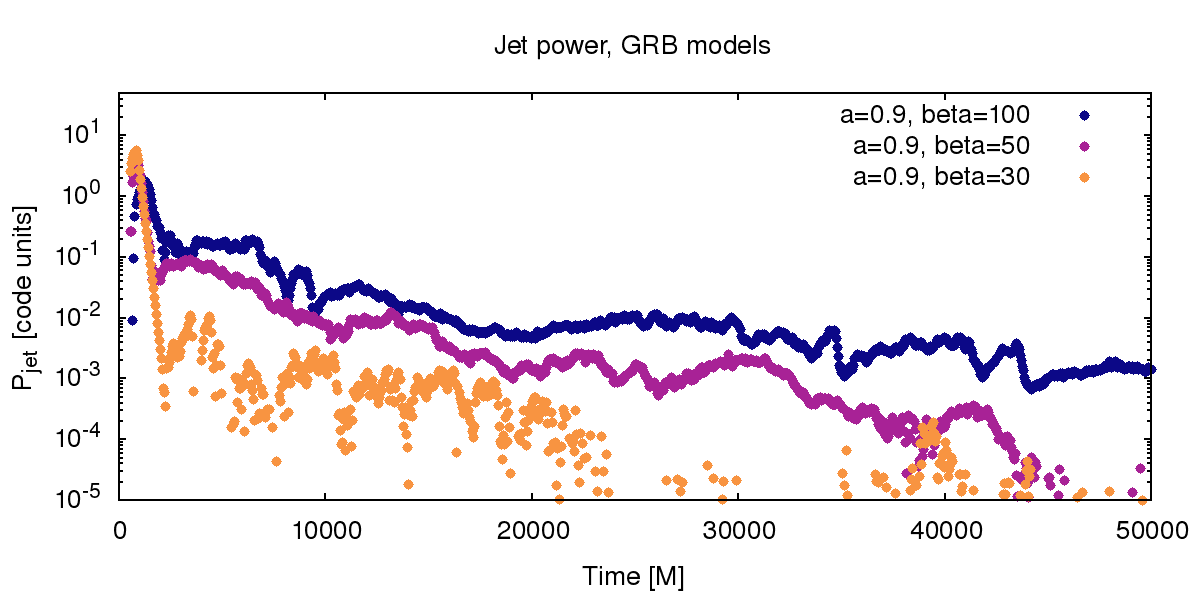}
    \includegraphics[width=0.45\textwidth]{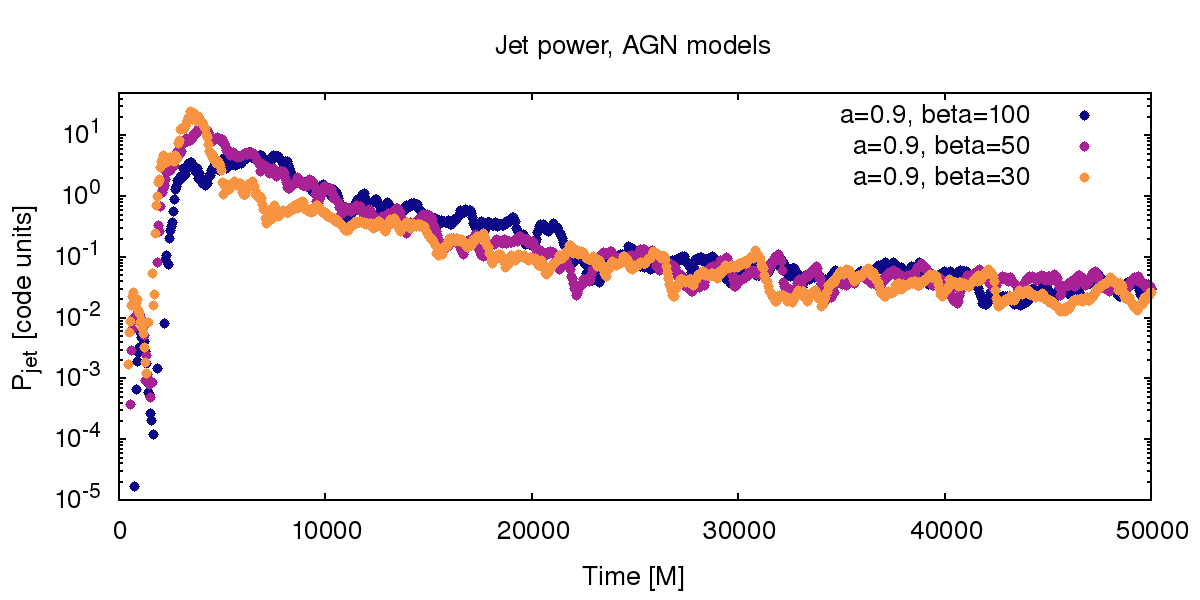} 
    \caption{Power of the Blandford-Znajek process, calculated from Eq. \ref{eq:power}, as a function of time. We present models GRB-HS and AGN-HS, shown in the top and bottom panels, respectively. Three colors denote different initial disk $\beta$-parameters, shown with blue, red, and green points for $\beta=100$, 50, and 30, respectively.}
    \label{fig:pjet}
\end{figure}

  In Figure \ref{fig:pjet}, we show the jet power as a function of time for GRB and AGN models. We chose the simulations with black hole spin of $a=0.9$ in both plots and we show the power (in code units) for various degrees of disk magnetization.
  We find that strong powerful jets are emitted in all AGN models with high black hole spin and that the jet power does not depend very much on the initial disk magnetization. For smaller $\beta$, the jet power evolves slightly faster at the beginning of the simulation to reach its equilibrium value. The power does not drop much until the end of the runs, that is, until time 50,000~$t_{g}$.
  In our GRB models, the initial power of the jet is smaller (in code units) than in the AGN case and it also drops by at least  three to four orders of magnitude by the end of the runs. For smallest initial $\beta$-parameter, the jet is quenched at the mid-time of simulation and its power drops to zero. The same occurs with the jet emitted in $\beta=50$ case, albeit at the end of the run. This behavior is correlated with the decrease of the dimensionless magnetic flux, $\phi_{BH}$ (cf. Fig. \ref{fig:timedep1a2a}).

The jet is driven by toroidal magnetic fields that operate in the accretion disks.
As shown by \citet{Mizuta2018}, the Alvfenic pulses in the jets are driven by the fluctuating toroidal field in the disk equatorial plane.
 Their toroidal component dominates over the poloidal field as a result of the disk's
  differential rotation.
Here, we plot the ratio of the toroidal to poloidal magnetic field components for two simulations, with both high spin and zero spin.

\begin{figure}
\centering
    \includegraphics[width=0.45\textwidth]{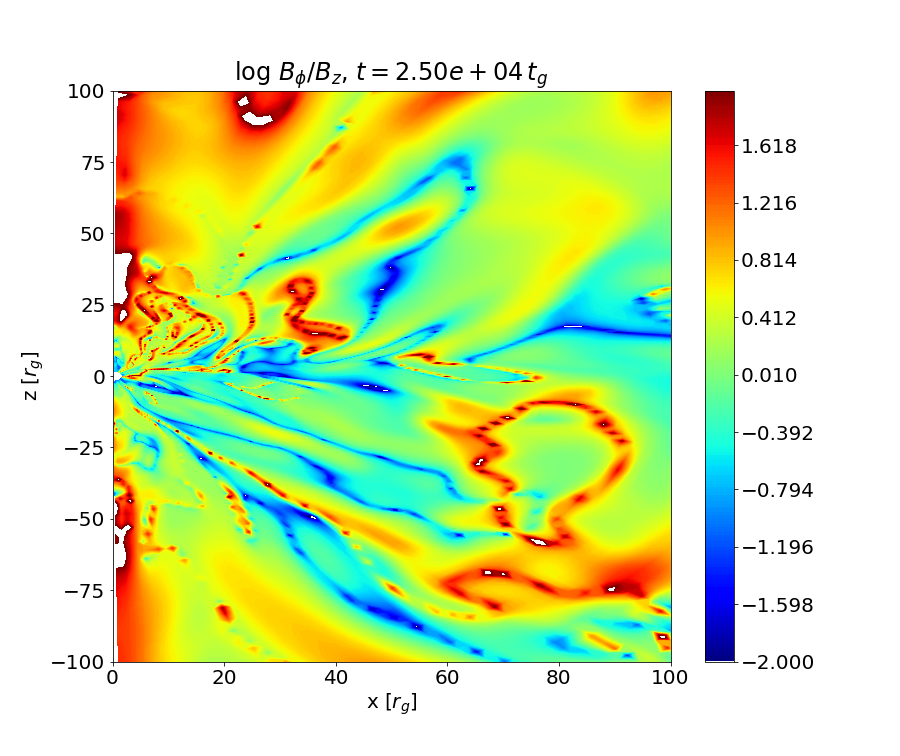}
    \includegraphics[width=0.45\textwidth]{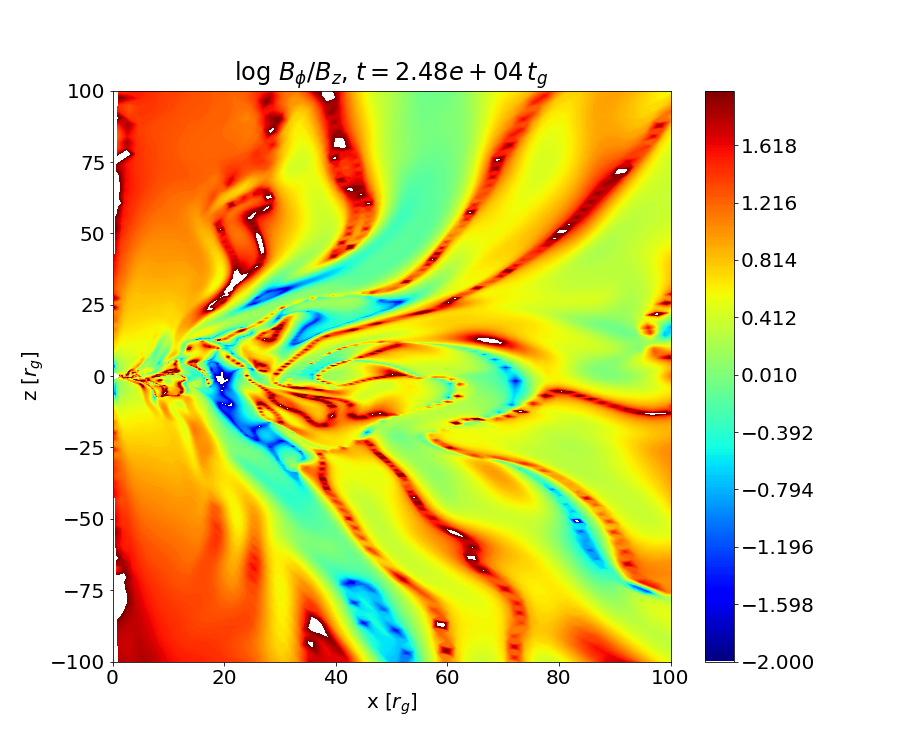} 
    \caption{Ratio of the toroidal to poloidal magnetic fields, ${B_{\phi}/B_{z}}$, for models AGN-NS50 and AGN-HS50, shown in the top and bottom panels, respectively. Color maps are in logarithmic scale}
    \label{fig:bphibz}
\end{figure}

The maps in Fig. \ref{fig:bphibz} are taken at the mid-time of the simulations, while both components of the magnetic field fluctuate over time. 
We  checked that (on average), the toroidal-to-poloidal magnetic field ratio in the jet is larger by two orders of magnitude in the model with high spin of the black hole  than for a non-spinning one. This toroidal field is enhanced by twisting due to the rotation of the central black hole. In addition, the $B_{\phi}/B_{z}$ ratio in this model
  is high at the disk equator up to the distance of $\sim 10 r_{g}$, because of the energy transfer from the rotating black hole to the torus via the magnetic coupling \citep{JaniukYuan2010}.

\begin{figure}
  \centering
 \includegraphics[width=0.24\textwidth]{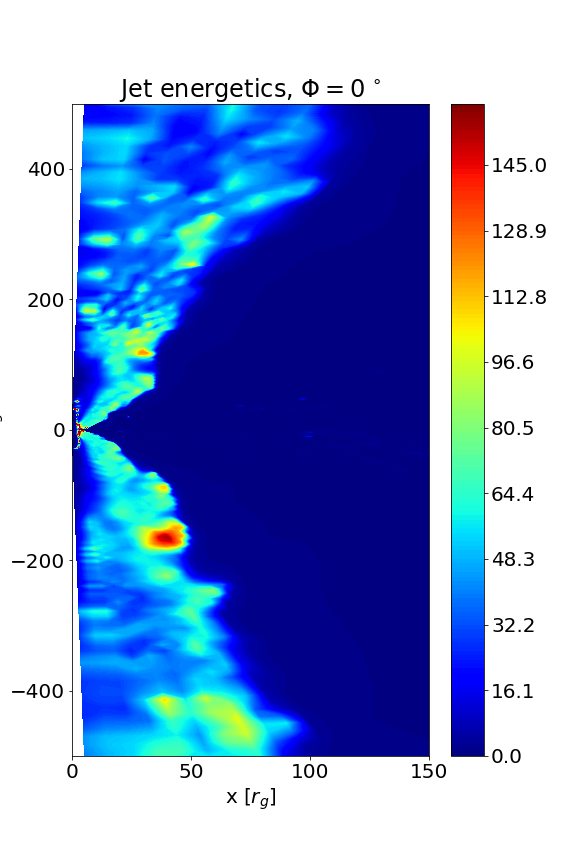} 
 \includegraphics[width=0.24\textwidth]{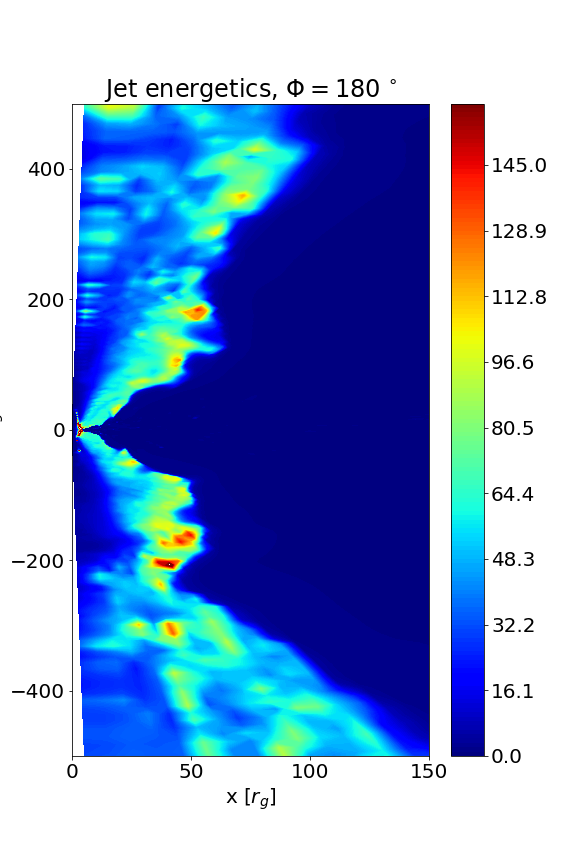} 
 \includegraphics[width=0.24\textwidth]{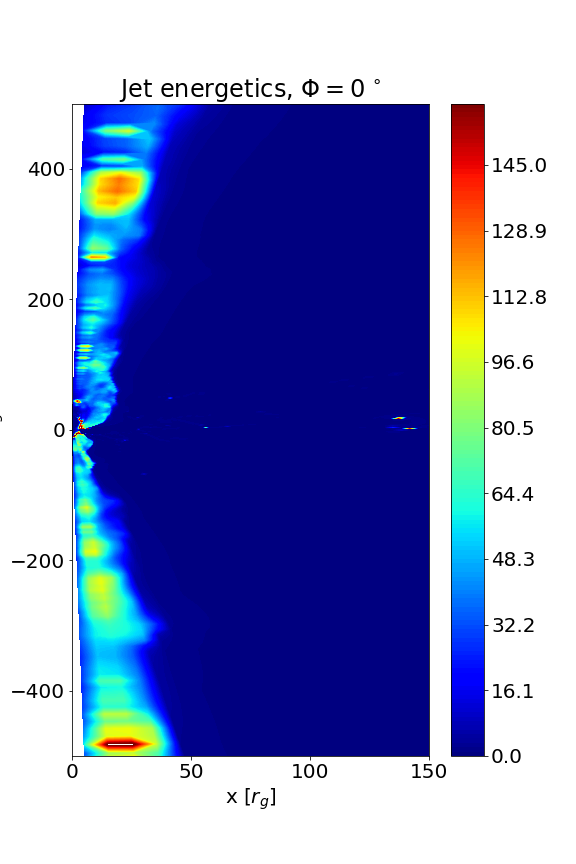} 
 \includegraphics[width=0.24\textwidth]{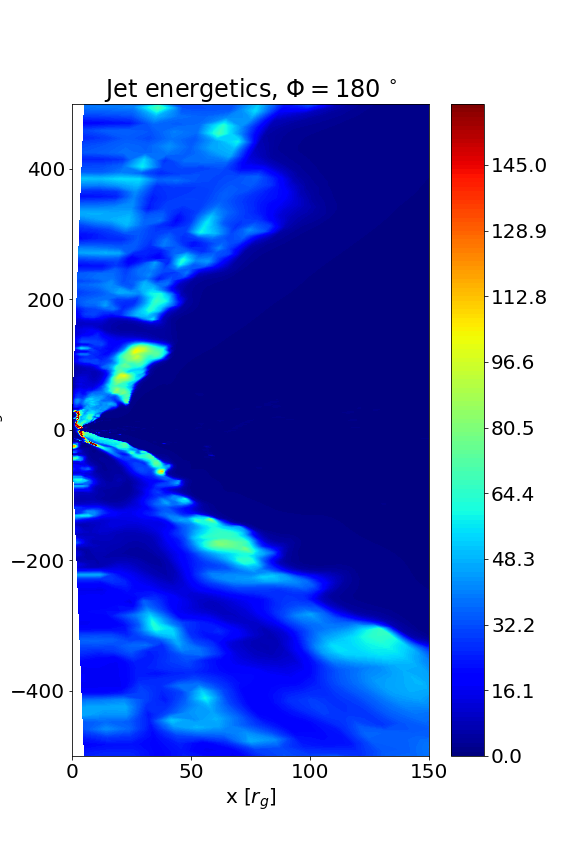} 
  \caption{Profiles of jet energetics, defined by $\mu$-parameter (see Eq. \ref{eq:jet_mu}) for two GRB models with high spin, $a=0.9$, and $\beta=100$ or $\beta=50$, on the two top, and the two bottom panels, respectively. The maps show profiles taken at two opposite azimuthal angles, $\Phi=0$ and $\Phi=\pi$. All snapshots are taken at a time of $t=2.5 \times 10^{4} t_{g}$}
  \label{fig:jetenerg}
  \end{figure}

In Figure \ref{fig:jetenerg} we plot the jets energetics structure, according to the definition:
\begin{equation}
    \mu = - {T^r_t \over \rho u^r}
\label{eq:jet_mu}
,\end{equation}
which is equivalent to the total (Poynting and thermal) energy of the jet, which can be transformed to the bulk kinetic content.
As can be seen from the plot, the jets are not axisymmetric, due to the central engine structure, and not uniform. The blobs of energy expanding with time lead to the jets variability (cf. \cite{Janiuk2021}). The plots compare shape of the jet base for two GRB models, with $\beta=100$ and $\beta=50$.
 We checked the trends also for $\beta=30$ and models with smaller black hole spins.
 The main relation between $\beta$ and the jet opening angle is that for larger  gas-to-magnetic pressure ratio, the jets seem to be more symmetric, while for smaller $\beta,$ the asymmetric shape makes the jet narrower at one viewing angle, while becoming broader at the opposite. The maximum energetics parameter depends mostly on black hole spin.
The azimuthally averaged jet profiles might appear more uniform \citep{Kati2019}.

\subsubsection{Mass outflow}

  We also checked the mass outflow through the outer boundary.
  It is very small at the beginning of the simulation, when the initial condition of the pressure equilibrium torus is being relaxed. However, after a time of about 20000 $t_{g}$, the outflow rate value is larger and some mass is left in the system.
  At the end of the simulation, the outflow rate is again small, below $10^{-3} - 10^{-7}$ (in code units).
We noticed a general trend that both higher black hole spin and higher magnetic pressure result in more powerful outflows and the maximal mass loss rate during the simulation scales with these parameters.

  We found that the mass loss rate is somewhat larger for the GRB models, while it is
  always negligibly small
  for our AGN models. The latter result seems in contradiction to the large mass outflow rates reported by \cite{2014MNRAS.441.3177M}. In their simulations, the radiatively driven outflows were more powerful, however, the difference might also be attributed to a different value of the outer radius in the simulations (our outer radius is five times larger).
  The total mass lost during the simulation via the unbound wind outflow in our models is given in Table \ref{tab:in}.

\subsubsection{Resolution of the simulations}

\begin{figure*}
\begin{tabular}{ccc}
  \hspace{-10mm}\includegraphics[width=0.59\textwidth]{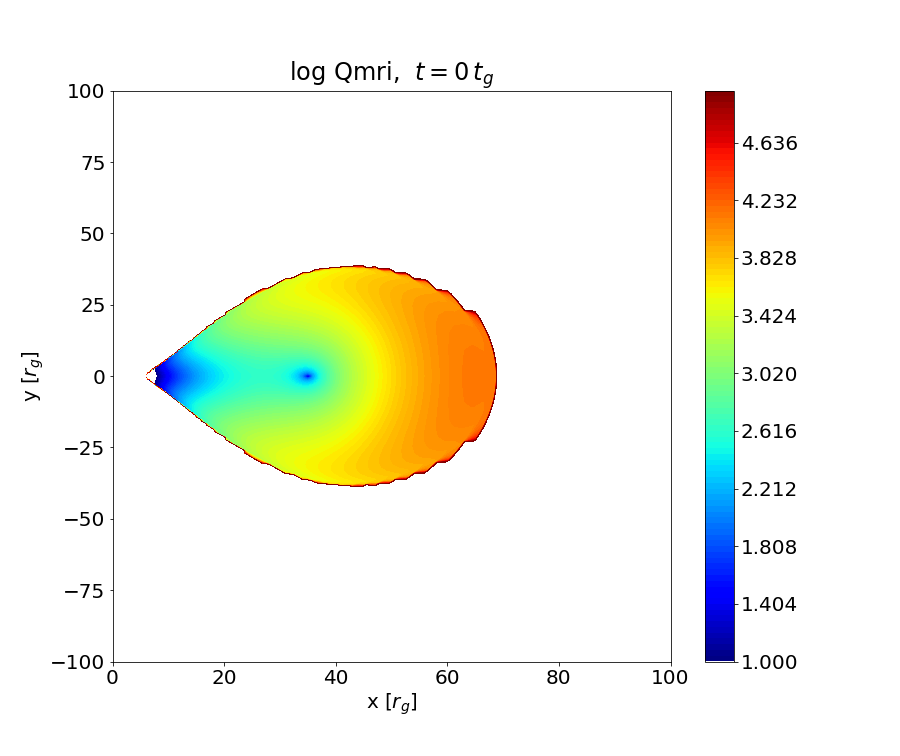} &
  \hspace{-10mm}\includegraphics[width=0.59\textwidth]{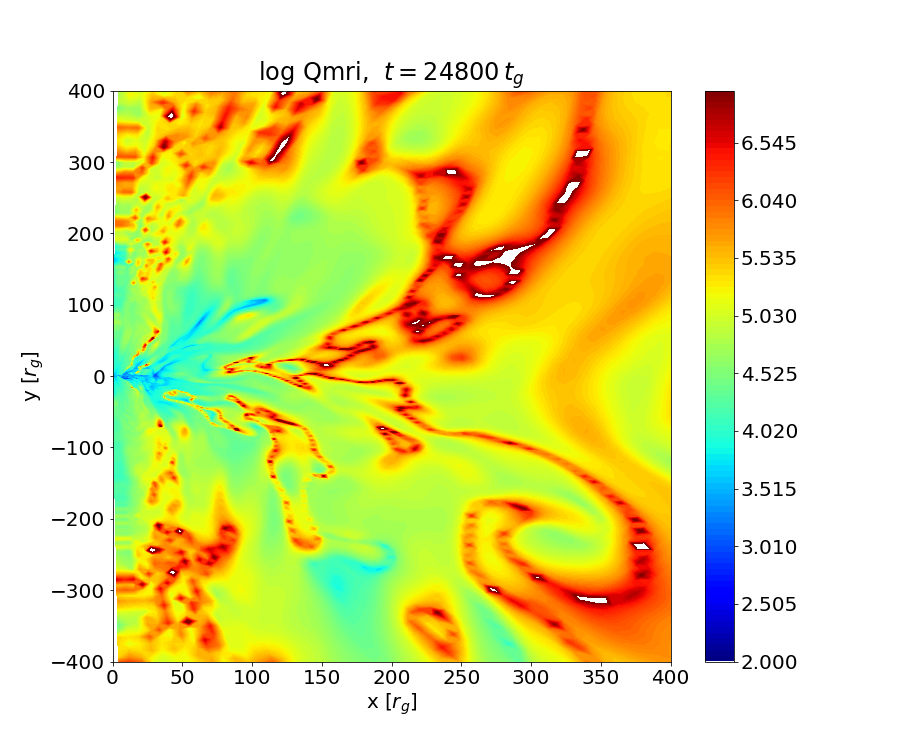} &
\\
  \hspace{-12mm}\includegraphics[width=0.59\textwidth]{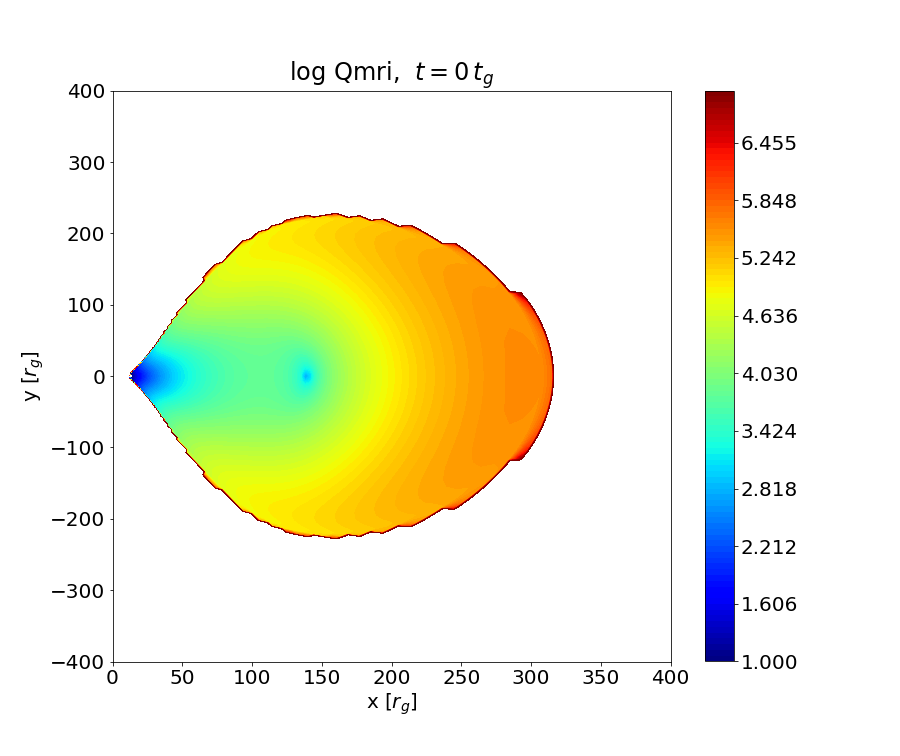}  &
  \hspace{-12mm}\includegraphics[width=0.59\textwidth]{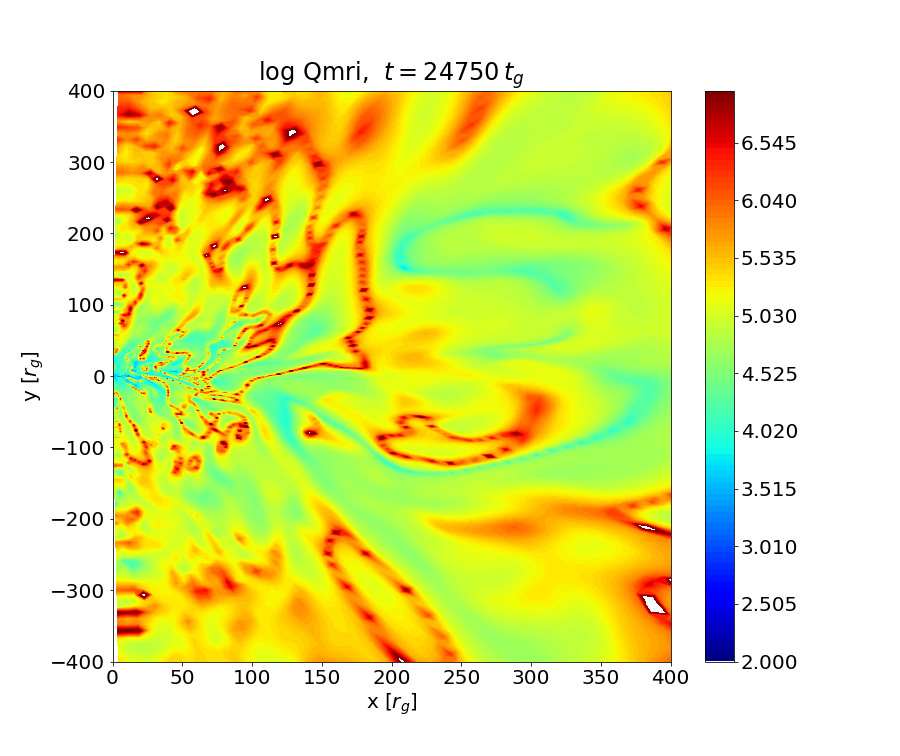}  &
\end{tabular}
\caption{Resolution of MRI instability, in the initial conditions and evolved state. The models are $GRB-HS100$ (top row) and $AGN-HS100$ (bottom row).
  The color maps are in logarithmic scale and are taken at $t=0$ (left) and
  at the intermediate time of the simulation, at $t \sim 25,000 M$ (right).
 }
    \label{fig:qmri}
\end{figure*}

The resolution of our simulations is moderate and would not be sufficient for the detailed studies of magnetic field reconnections.
However, we checked that the number of cells is adequate to resolve the magnetorotational instability (MRI). In Figure \ref{fig:qmri}, we plot the quantity $Q_{MRI}$, defined as the number of grid cells in polar direction, per wavelength of the fastest growing mode (cf. \cite{Siegel2018}):
\begin{equation}
\lambda_{\rm MRI} = { 2 \pi \over \Omega} {b \over \sqrt {4\pi \rho h + b^{2}}}
  .\end{equation}
The MRI is well resolved, when $Q_{\rm MRI}={\lambda_{\rm MRI} \over {\Delta\theta}}$ is larger than 10. This requirement is satisfied in our initial conditions, even in models with a weak magnetic field, as shown in the figure.

We also note that the results of the simulation might be somewhat sensitive to the density and energy floors adopted to set their minimum values handled by the numerical scheme. The floor values in our calculations were set to $10^{-8}$ and $10^{-10}$ for the minimum density and internal energy (in code units), respectively. These values, however, mainly influence the simulation results at the initial time, when the density contrasts are high.
As shown by \citet{Sap2019ApJ}, the time-averaged distributions, for instance, those of the jet energetics, are saturated at the same level for different floor values.

\section{Discussion}

We studied fully 3D general relativistic numerical simulations of accretion flows that lead to a magnetically arrested (MAD) state and that can describe the active centers of quasars or engines of gamma ray bursts. Our simulations adopt an adiabatic equation of state that neglects the microphysics effects and that are performed in a dimensionless setup. Hence, with the scaling of the black hole mass and the torus density, this universal setup can be used for both supermassive and stellar mass accreting black holes. In particular, the timescales of flares and the time variability of energy injection to the jets, are scaled with geometric time, $t_{g}=G M_{BH} c^{-3}$. Therefore, they can be directly compared to both
 supermassive and stellar mass accreting black holes, in AGN and GRBs.
The timescales corresponding to the
flaring activity of blazars, such as that observed in Mrk 421 \citep{Chatterjee2021} or in Mrk 501 \citep{Giommi2021}, are less than 1 day. Similarly, the variability timescales in Sgr A$\star$ can reveal about 1 flare per day, as reported by \citep{neilsen2013} for the \textit{Chandra} observations, or up to 2.5 flares per day on average, found in \textit{Chandra} and \textit{XMM-Newton} data, as reported by \cite{ponti2015}.
The observed variability of gamma ray emission in GRBs down to millisecond timescales is correlated with the burst duration \citep{McLachlan2013}.
It has also been found that a correlation exists between the burst variability and its intrinsic luminosity, suggesting a Cepheid-like relation for these sources, for a sample observed mainly by \textit{BATSE} \citep{Reichart2001}.
The absolute variability timescales are related to the black hole mass and this an approach was already taken in our previous study \citep{Janiuk2021} to explain and observed correlation between the jet Lorentz factor and minimum variability timescale  across the mass scale \citep{wu2016}.

The variability of prompt emission of GRBs is related to both variability of the central engine itself and to the interaction of the jet with
 ambient medium, such as disk wind, the post-merger ejecta in a
  short GRB \citep{Murguia_Berthier2021}, or with the
progenitor star envelope in a long GRB case \citep{Morsony}.
The latter may introduce an observed variability even if the central engine fluctuations are very slight.

In the current simulations, we focus on the engine, and jet base variability, as driven by magneto-rotational instability, or interchange instabilities in MAD disks.
The jet propagation inside the star is a complex non-linear process, where the turbulence develops and energy may be dissipated in the form of shocks. As this aspect is ignored in our simulations, it basically represents the situation  where the engine activity timescale is much longer than the jet propagation timescale.

Our analysis shows that the accretion rate variations at the inner edge of the disk may be responsible for the observed variability of the sources. The PDS\ have slopes between
1.67-1.97
for the highly spinning black hole engine
in the GRB models, while the PDS slopes for moderate black hole spins are
in the range
1.64-1.8.
Our simulated PDS spectra are comparable the observed variability of GRBs. The calculated PDS slopes given in Table \ref{tab:pds_results} are roughly consistent with the canonical value of power-law PDS with $\alpha=5/3$ reported by \cite{Beloborodov2000} for the bolometric light curves of long-duration GRBs.
More recent observations show that the mean value of PDS slope found by \cite{Ukwatta2011} was equal to $\alpha=1.4 \pm 0.6$, while the
  variability studies based on \textit{Swift} observations of individual GRBs presented by \cite{Guidorzi2016} revealed a much broader range of PDS slopes, namely, $\alpha=1.3-3.2$. These authors also found that the number of pulses in the GRB light curve is strongly anticorrelated with the PDS slope. We note, however, that most of the variability studies have been based on the long GRB data, which are easier to analyze thanks to their long time series. The outlier short GRB source reported in \cite{Dichiara2016} was found to have a very steep PDS slope of about $\alpha=3.5$, while \cite{Dichiara2013} found the PDS slope for the sample of 44 short GRBs to be between 1.4 and 2.5.

  The variability of supermassive accreting black holes measured in the X-ray, radio, and NIR bands shows the variations of synchrotron radiation. As found by \cite{Witzel2018}, the NIR light curve variations during the low flux density phase of Sgr A$\star$ are best fitted by PDS with a slope of $\alpha=1.6$. We notice that this value is in agreement with our model results for the \textit{AGN-HS} simulations, namely, the required spin of the black hole would be $a=0.9$. If the black hole in Sgr A$\star$ is a non-spinning one, then the only simulation consistent with the observed variability of this source is the \textit{AGN-NS30} model, namely, the one with largest magnetization.
  
The minimum variability timescale inferred from our models is between 25-30 and 70-90 gravitational timescales, depending on the black hole spin. For the supermassive black hole in M87, the timescale of 30 $t_{g}$ is equal to about 8.5 days (cf. Tab. \ref{tab:conversion}). Hence, our AGN models with highly spinning black hole may account for the flaring activity of this source, as observed in the very high energies \citep{Aharonian2003, Acciari2009}.

The jet may interact with the disk winds, which are expected to accompany the GRB prompt phase.
Such interaction can occur only
when the wind mass loss rate is large.
The total mass loss rate in our models is given in Table \ref{tab:in}. We note that this is only a lower limit, as the simulation duration is not long enough to account for the mass loss calculation through the outer boundary. Nevertheless, we show that there should be a correlation between mass loss rate and disk magnetization as well as with the black hole spin.
Therefore, we expect that the jet can be partially or completely choked, in the most magnetized scenario.  The results displayed in Figure \ref{fig:pjet} confirm this effect. 
In the models with $\beta=30$, the wind density is high, while the jet power is smaller than for the higher $\beta$ parameter. The jets launched from a moderately spinning black hole, however, seem to have more chance to successfully break through the wind, than those launched from highly spinning black holes. This is because
the wind density is small.
 The interaction with disk winds should impose an additional variability in the observed emission, as the outflow propagation within a denser medium will
  produce irregularities of the Rayleigh-Taylor instability. As shown by
\cite{Lazzati2021}, the jet's centroid oscillates around the axis of the system, due to inhomogeneities encountered in the propagation, and the jet slows down on a very short timescale.

In the current 3D simulation, we use the simplified microphysics treatment in the case of GRB disks. 
This is done in part to allow for a unified modeling of both GRB and AGN types of sources, but also partially reflects our computational limitations.
The equation of state of dense matter and neutrino cooling effects have been implemented into our numerical scheme, \citep{Janiuk2019}; however, because of the computational resource limitations, they effectively work in 2D geometry. The full 3D numerical simulation of the magnetically arrested disk in the GRB engine has been postponed to a future work. At this stage, we speculate that the quantitative differences in the results will be important mostly for the physics of the unbound outflows, namely, winds that are affected by both neutrinos and magnetically driven acceleration. In this way, the winds may be more dense and powerful if the neutrino-driven mechanism supports or substitutes the magnetically driven wind \citep{Murguia_Berthier2021}.

In the initial setup, we adopted the poloidal configuration of magnetic fields, with the field lines that follow the iso-contours of constant density. Our choice is similar to that of other authors. Initially, \cite{McKinney_2012} noticed that initially
toroidally-dominated models lead to only patches of magnetic
flux and transient outflows or weak jets (see also \cite{Beckwith_2008, Paschalidis_Ruiz_Shapiro_2015}).
 More recent works have shown that jets form even from purely toroidal magnetic field initial conditions \citep{Christie2019, Liska2020}.
Due to the dynamo mechanism, the toroidal component of magnetic field develops over time during the simulation and exceeds the vertical field component. The specific choice of the magnetic field configuration also assumes  that the magnetic field is additionally scaled with the distance to allow for  the magnetic pressure to build up as the simulation continues.

The assumption of the same magnetic fields configuration for both AGN and GRB models is motivated by some subtle arguments. In principle, there is no obvious reason to believe that magnetic fields geometries should differ between AGN and stellar mass-accreting black holes, while both types of systems are able to launch relativistic jets on the cost of black hole rotation, mediated by magnetic fields \citep{DaviesRev}. On the other hand, the GRBs jets are systematically stronger, while the magnetic fields amplification may originate not only from the effective accretion of a large-scale magnetic field to the inner regions, but may also be subject to the dynamo process acting during the stellar collapse or neutron star merger. Recent numerical simulations performed in this context  have not been conclusive, however, about the type of configuration that should be favored 
\citep{Ponce2014, Ruiz2020, Shibata2021}.

Our simulations are addressed to both AGN and GRB engines, while the calculations are limited to the innermost regions of the accretion flow and spatial scales close to the base of the jets. Models were initiated from a hydrostatic torus solution and evolution was triggered with a random perturbation of the thermal pressure, which allows for the formation of non-axisymmetric modes of Rayleigh-Taylor instabilities. These are important for allowing the material to overpass the magnetic barrier and continue accretion when large magnetic flux is accumulated on the black hole horizon.
The magnetic fields are mainly responsible for the transport
of the angular momentum, which enables accretion, but also 
drives the disk outflows.
The time-averaged radial magnetic flux on the black hole horizon, measured in our simulations, is on the order of $(1.5-2.5)\cdot 10^{28}$ up to $1.6\cdot 10^{29}$ G cm$^{2}$ for the GRB models, depending on the adopted scaling (cf. Table \ref{tab:conversion}). This range is consistent with the amount of oriented magnetic flux in supernova progenitors \citep{Heger2005, obergaulinger} and with the field strengths in coalescing binaries of neutron stars \citep{Kuan2021}.
  Magnetic flux scaling for the supermassive black hole in our Galaxy center gives
  the range of $5-7.5\cdot 10^{-7}$G pc$^{2}$, which can be supported by both the interstellar medium and massive magnetized O-type stars feeding the galactic nucleus (cf. \cite{Witzel2021, Hubrig2020}).

The initial condition adopted for the density structure, namely, the equilibrium torus, affects the simulations at early times, while it needs to be relaxed via the action of magnetic fields. We overcome the drawback of such an artificial initial condition by running a simulation for a sufficiently long time, then we follow the time evolution of the system for long time, up to $50,000$ $t_{g}$. We note that an alternative approach would be to use an initial setup resulting from an evolved postmerger state provided by previous numerical relativity simulations \citep{PhysRevD.97.083014}.
After the formation of the large-scale open field lines, the Blandford-Znajek process extracts the rotational energy of the BH. This process is common to both AGN and GRB engines, while the power of jets obtained for these two types of sources differs. The magnetic flux available for accretion in GRBs is on the order of $10^{-9}$ G pc$^{2}$, and for active galaxies, it is a few orders of magnitude larger. Nevertheless, the density of accreting plasma is much larger in GRB engines, therefore, due to the compactness of the system the total power available for a jet is larger in these objects.
In addition, the neutrino annihilation can be a complementary process to power the GRB jets \citep{Mochkovitch_et_al_1993, Aloy_et_al_2005, Janiuk2013, Liu2015, Janiuk2017}, but this feature should not change our main results.
It is possible, however, that the GRB jets are choked in dense wind if it is driven both magnetically and by neutrino heating. Such a simulation is currently beyond the scope of this paper, but it is planned in the future,  based on the formalism derived in \cite{Janiuk2019}.
Recent works on this topic, for instance, by \cite{Siegel2018, Fernandez2019}, have shown that powerful outflows emerge from BNS merger remnant disks and can be sites of r-process nucleosynthesis, while the bipolar jets in these simulations have only mildly relativistic velocities, with Lorentz factors of 1-10.

  Finally, we note that in our models, the adiabatic index of $\gamma=4/3$ is used for both AGN and GRB scenarios. In fact, the ideal gas law with $\gamma=5/3$ could be used to describe low-luminosity AGN, while the relativistically hot gas with $\gamma=4/3$ is more likely to describe GRB engines which are cooled by neutrino emission. In super-Eddington accretion disks, the radiative cooling is not efficient either, as lot of the dissipated energy is advected toward the black hole \citep{2019MNRAS.489..524K}.
  These disks are geometrically thick and can readily be described by our simulations.
  The effect of changing the adiabatic index may influence the frequency and amplitude of accretion rate oscillations, especially if the flow undergoes a shock compression \citep{Palit2019MNRAS}.

  \section{Conclusions}

  We performed a series of full 3D general relativistic numerical simulations of magnetically arrested disk and jets launched from these central engines.
  We found that the system evolution is governed by the physical parameters of the engine, such as the black hole spin, and disk size, as well as disk magnetization, and we applied our scenarios to typical types of sources
  for both AGN and GRB classes.
  We found that the MAD scenario is applicable to AGN engines and supports persistent
  jet emissions. The variability pattern is, however, consistent with currently available observations only if we assume high black hole spins.
  We also find that while the MAD scenario can be applied to GRBs, our jet power estimates are quite large and cannot explain faint sources.
  Still, these simulations give a variability pattern that is roughly consistent with GRB power density spectra.
  Finally, we found that in some cases, strong magnetic fields may lead 
to jet quenching and this effect is found to be important mainly for
gamma ray burst jets. Finally we speculate that this may be related to the strength of magnetically driven winds from the GRB engines.

\begin{acknowledgements}
We thank Bozena Czerny, Marek Sikora and Krzysztof Nalewajko for helpful discussions. We also thank Kostas Sapountzis, for help in running simulations, and preparing some plotting scripts in Jupyter Notebook.
This work was supported in part by the grant
2019/35/B/ST9/04000 from the Polish National Science Center.
We also acknowledge support from the Interdisciplinary Center for Mathematical Modeling of the Warsaw University, and the PL-Grid infrastructure, through the computational grant plggrb5.
\end{acknowledgements}

\bibliographystyle{aa}
\bibliography{paper_mad}

\begin{thebibliography}{94}
\expandafter\ifx\csname natexlab\endcsname\relax\def\natexlab#1{#1}\fi

\bibitem[{{Acciari} {et~al.}(2009){Acciari}, {Aliu}, {Arlen}, {Bautista},
  {Beilicke}, {Benbow}, {Bradbury}, {Buckley}, {Bugaev}, {Butt}, {Byrum},
  {Cannon}, {Celik}, {Cesarini}, {Chow}, {Ciupik}, {Cogan}, {Cui},
  {Dickherber}, {Fegan}, {Finley}, {Fortin}, {Fortson}, {Furniss}, {Gall},
  {Gillanders}, {Grube}, {Guenette}, {Gyuk}, {Hanna}, {Holder}, {Horan}, {Hui},
  {Humensky}, {Imran}, {Kaaret}, {Karlsson}, {Kieda}, {Kildea}, {Konopelko},
  {Krawczynski}, {Krennrich}, {Lang}, {LeBohec}, {Maier}, {McCann},
  {McCutcheon}, {Millis}, {Moriarty}, {Ong}, {Otte}, {Pandel}, {Perkins},
  {Petry}, {Pohl}, {Quinn}, {Ragan}, {Reyes}, {Reynolds}, {Roache}, {Roache},
  {Rose}, {Schroedter}, {Sembroski}, {Smith}, {Swordy}, {Theiling}, {Toner},
  {Varlotta}, {Vincent}, {Wakely}, {Ward}, {Weekes}, {Weinstein}, {Williams},
  {Wissel}, {Wood}, {Walker}, {Davies}, {Hardee}, {Junor}, {Ly}, {Aharonian},
  {Akhperjanian}, {Anton}, {Barres de Almeida}, {Bazer-Bachi}, {Becherini},
  {Behera}, {Bernl{\"o}hr}, {Bochow}, {Boisson}, {Bolmont}, {Borrel},
  {Brucker}, {Brun}, {Brun}, {B{\"u}hler}, {Bulik}, {B{\"u}sching},
  {Boutelier}, {Chadwick}, {Charbonnier}, {Chaves}, {Cheesebrough}, {Chounet},
  {Clapson}, {Coignet}, {Dalton}, {Daniel}, {Davids}, {Degrange}, {Deil},
  {Dickinson}, {Djannati-Ata{\"\i}}, {Domainko}, {Drury}, {Dubois}, {Dubus},
  {Dyks}, {Dyrda}, {Egberts}, {Emmanoulopoulos}, {Espigat}, {Farnier},
  {Feinstein}, {Fiasson}, {F{\"o}rster}, {Fontaine}, {F{\"u}{\ss}ling},
  {Gabici}, {Gallant}, {G{\'e}rard}, {Gerbig}, {Giebels}, {Glicenstein},
  {Gl{\"u}ck}, {Goret}, {G{\"o}hring}, {Hauser}, {Hauser}, {Heinz},
  {Heinzelmann}, {Henri}, {Hermann}, {Hinton}, {Hoffmann}, {Hofmann},
  {Holleran}, {Hoppe}, {Horns}, {Jacholkowska}, {de Jager}, {Jahn}, {Jung},
  {Katarzy{\'n}ski}, {Katz}, {Kaufmann}, {Kendziorra}, {Kerschhaggl},
  {Khangulyan}, {Kh{\'e}lifi}, {Keogh}, {Klu{\'z}niak}, {Kneiske}, {Komin},
  {Kosack}, {Lamanna}, {Lenain}, {Lohse}, {Marandon}, {Martin},
  {Martineau-Huynh}, {Marcowith}, {Maurin}, {McComb}, {Medina}, {Moderski},
  {Moulin}, {Naumann-Godo}, {de Naurois}, {Nedbal}, {Nekrassov}, {Nicholas},
  {Niemiec}, {Nolan}, {Ohm}, {Olive}, {O{\~n}a de Wilhelmi}, {Orford},
  {Ostrowski}, {Panter}, {Paz Arribas}, {Pedaletti}, {Pelletier}, {Petrucci},
  {Pita}, {P{\"u}hlhofer}, {Punch}, {Quirrenbach}, {Raubenheimer}, {Raue},
  {Rayner}, {Renaud}, {Rieger}, {Ripken}, {Rob}, {Rosier-Lees}, {Rowell},
  {Rudak}, {Rulten}, {Ruppel}, {Sahakian}, {Santangelo}, {Schlickeiser},
  {Sch{\"o}ck}, {Schr{\"o}der}, {Schwanke}, {Schwarzburg}, {Schwemmer},
  {Shalchi}, {Sikora}, {Skilton}, {Sol}, {Spangler}, {Stawarz}, {Steenkamp},
  {Stegmann}, {Stinzing}, {Superina}, {Szostek}, {Tam}, {Tavernet}, {Terrier},
  {Tibolla}, {Tluczykont}, {van Eldik}, {Vasileiadis}, {Venter}, {Venter},
  {Vialle}, {Vincent}, {Vivier}, {V{\"o}lk}, {Volpe}, {Wagner}, {Ward},
  {Zdziarski}, {Zech}, {Anderhub}, {Antonelli}, {Antoranz}, {Backes},
  {Baixeras}, {Balestra}, {Barrio}, {Bastieri}, {Becerra Gonz{\'a}lez},
  {Becker}, {Bednarek}, {Berger}, {Bernardini}, {Biland}, {Bock}, {Bonnoli},
  {Bordas}, {Tridon}, {Bosch-Ramon}, {Bose}, {Braun}, {Bretz}, {Britvitch},
  {Camara}, {Carmona}, {Commichau}, {Contreras}, {Cortina}, {Costado},
  {Covino}, {Curtef}, {Dazzi}, {De Angelis}, {de Cea del Pozo}, {Delgado
  Mendez}, {De los Reyes}, {De Lotto}, {De Maria}, {De Sabata}, {Dominguez},
  {Dorner}, {Doro}, {Elsaesser}, {Errando}, {Ferenc}, {Fern{\'a}ndez}, {Firpo},
  {Fonseca}, {Font}, {Galante}, {Garc{\'\i}a L{\'o}pez}, {Garczarczyk}, {Gaug},
  {Goebel}, {Hadasch}, {Hayashida}, {Herrero}, {Hildebrand},
  {H{\"o}hne-M{\"o}nch}, {Hose}, {Hsu}, {Jogler}, {Kranich}, {La Barbera},
  {Laille}, {Leonardo}, {Lindfors}, {Lombardi}, {Longo}, {L{\'o}pez}, {Lorenz},
  {Majumdar}, {Maneva}, {Mankuzhiyil}, {Mannheim}, {Maraschi}, {Mariotti},
  {Mart{\'\i}nez}, {Mazin}, {Meucci}, {Miranda}, {Mirzoyan}, {Miyamoto},
  {Mold{\'o}n}, {Moles}, {Moralejo}, {Nieto}, {Nilsson}, {Ninkovic}, {Oya},
  {Paoletti}, {Paredes}, {Pasanen}, {Pascoli}, {Pauss}, {Pegna},
  {Perez-Torres}, {Persic}, {Peruzzo}, {Prada}, {Prandini}, {Puchades},
  {Reichardt}, {Rhode}, {Rib{\'o}}, {Rico}, {Rissi}, {Robert}, {R{\"u}gamer},
  {Saggion}, {Saito}, {Salvati}, {Sanchez-Conde}, {Satalecka}, {Scalzotto},
  {Scapin}, {Schweizer}, {Shayduk}, {Shore}, {Sidro}, {Sierpowska-Bartosik},
  {Sillanp{\"a}{\"a}}, {Sitarek}, {Sobczynska}, {Spanier}, {Stamerra}, {Stark},
  {Takalo}, {Tavecchio}, {Temnikov}, {Tescaro}, {Teshima}, {Torres}, {Turini},
  {Vankov}, {Wagner}, {Zabalza}, {Zandanel}, {Zanin}, {Zapatero}, {VERITAS
  Collaboration}, {VLBA 43 GHz M87 Monitoring Team}, {H.~E.~S.~S.
  Collaboration}, \& {MAGIC Collaboration}}]{Acciari2009}
{Acciari}, V.~A., {Aliu}, E., {Arlen}, T., {et~al.} 2009, Science, 325, 444

\bibitem[{{Aharonian} {et~al.}(2003){Aharonian}, {Akhperjanian}, {Beilicke},
  {Bernl{\"o}hr}, {B{\"o}rst}, {Bojahr}, {Bolz}, {Coarasa}, {Contreras},
  {Cortina}, {Denninghoff}, {Fonseca}, {Girma}, {G{\"o}tting}, {Heinzelmann},
  {Hermann}, {Heusler}, {Hofmann}, {Horns}, {Jung}, {Kankanyan}, {Kestel},
  {Kohnle}, {Konopelko}, {Kornmeyer}, {Kranich}, {Lampeitl}, {Lopez}, {Lorenz},
  {Lucarelli}, {Mang}, {Meyer}, {Mirzoyan}, {Moralejo}, {Ona-Wilhelmi},
  {Panter}, {Plyasheshnikov}, {P{\"u}hlhofer}, {de los Reyes}, {Rhode},
  {Ripken}, {Rowell}, {Sahakian}, {Samorski}, {Schilling}, {Siems},
  {Sobzynska}, {Stamm}, {Tluczykont}, {Vitale}, {V{\"o}lk}, {Wiedner}, \&
  {Wittek}}]{Aharonian2003}
{Aharonian}, F., {Akhperjanian}, A., {Beilicke}, M., {et~al.} 2003, \aap, 403,
  L1

\bibitem[{Aloy {et~al.}(2005)Aloy, Janka, \& M{\"u}ller}]{Aloy_et_al_2005}
Aloy, M.~A., Janka, H.-T., \& M{\"u}ller, E. 2005, \aap, 436, 273

\bibitem[{{Avara} {et~al.}(2016){Avara}, {McKinney}, \& {Reynolds}}]{Avara2016}
{Avara}, M.~J., {McKinney}, J.~C., \& {Reynolds}, C.~S. 2016, \mnras, 462, 636

\bibitem[{{Baglio} {et~al.}(2018){Baglio}, {Russell}, {Casella}, {Noori},
  {Yazeedi}, {Belloni}, {Buckley}, {Cadolle Bel}, {Ceccobello}, {Corbel}, {Coti
  Zelati}, {D{\'\i}az Trigo}, {Fender}, {Gallo}, {Gandhi}, {Homan}, {Koljonen},
  {Lewis}, {Maccarone}, {Malzac}, {Markoff}, {Miller-Jones}, {O'Brien},
  {Russell}, {Saikia}, {Shahbaz}, {Sivakoff}, {Soria}, {Testa}, {Tetarenko},
  {van den Ancker}, \& {Vincentelli}}]{Baglio2018}
{Baglio}, M.~C., {Russell}, D.~M., {Casella}, P., {et~al.} 2018, \apj, 867, 114

\bibitem[{Balbus \& Hawley(1991)}]{Balbus_Hawley_1991}
Balbus, S.~A. \& Hawley, J.~F. 1991, \apj, 376, 214

\bibitem[{Beckwith {et~al.}(2008)Beckwith, Hawley, \& Krolik}]{Beckwith_2008}
Beckwith, K., Hawley, J.~F., \& Krolik, J.~H. 2008, \apj, 678, 1180

\bibitem[{{Beloborodov} {et~al.}(2000){Beloborodov}, {Stern}, \&
  {Svensson}}]{Beloborodov2000}
{Beloborodov}, A.~M., {Stern}, B.~E., \& {Svensson}, R. 2000, \apj, 535, 158

\bibitem[{Blandford \& Znajek(1977)}]{Blandoford_Znajek_1977}
Blandford, R.~D. \& Znajek, R.~L. 1977, \mnras, 179, 433

\bibitem[{{Broderick} {et~al.}(2011){Broderick}, {Fish}, {Doeleman}, \&
  {Loeb}}]{2011ApJ...735..110B}
{Broderick}, A.~E., {Fish}, V.~L., {Doeleman}, S.~S., \& {Loeb}, A. 2011, \apj,
  735, 110

\bibitem[{{Chatterjee} {et~al.}(2021){Chatterjee}, {Das}, {Khasnovis}, {Ghosh},
  {Kumari}, {Naik}, {Larionov}, {Grishina}, {Kopatskaya}, {Larionova},
  {Nikiforova}, {Morozov}, {Savchenko}, {Troitskaya}, {Troitsky}, \&
  {Vasilyev}}]{Chatterjee2021}
{Chatterjee}, R., {Das}, S., {Khasnovis}, A., {et~al.} 2021, Journal of
  Astrophysics and Astronomy, 42, 80

\bibitem[{{Christie} {et~al.}(2019){Christie}, {Lalakos}, {Tchekhovskoy},
  {Fern{\'a}ndez}, {Foucart}, {Quataert}, \& {Kasen}}]{Christie2019}
{Christie}, I.~M., {Lalakos}, A., {Tchekhovskoy}, A., {et~al.} 2019, \mnras,
  490, 4811

\bibitem[{{Corsi} \& {Lazzati}(2021)}]{Corsi2021}
{Corsi}, A. \& {Lazzati}, D. 2021, \nar, 92, 101614

\bibitem[{{Cuadra} {et~al.}(2005){Cuadra}, {Nayakshin}, {Springel}, \& {Di
  Matteo}}]{2005MNRAS.360L..55C}
{Cuadra}, J., {Nayakshin}, S., {Springel}, V., \& {Di Matteo}, T. 2005, \mnras,
  360, L55

\bibitem[{{Czerny} {et~al.}(2016){Czerny}, {Du}, {Wang}, \&
  {Karas}}]{Czerny2016}
{Czerny}, B., {Du}, P., {Wang}, J.-M., \& {Karas}, V. 2016, \apj, 832, 15

\bibitem[{{Davis} \& {Tchekhovskoy}(2020)}]{DaviesRev}
{Davis}, S.~W. \& {Tchekhovskoy}, A. 2020, \araa, 58, 407

\bibitem[{{Dexter} {et~al.}(2014){Dexter}, {McKinney}, {Markoff}, \&
  {Tchekhovskoy}}]{dexter}
{Dexter}, J., {McKinney}, J.~C., {Markoff}, S., \& {Tchekhovskoy}, A. 2014,
  \mnras, 440, 2185

\bibitem[{{Dichiara} {et~al.}(2016){Dichiara}, {Guidorzi}, {Amati}, {Frontera},
  \& {Margutti}}]{Dichiara2016}
{Dichiara}, S., {Guidorzi}, C., {Amati}, L., {Frontera}, F., \& {Margutti}, R.
  2016, \aap, 589, A97

\bibitem[{{Dichiara} {et~al.}(2013){Dichiara}, {Guidorzi}, {Frontera}, \&
  {Amati}}]{Dichiara2013}
{Dichiara}, S., {Guidorzi}, C., {Frontera}, F., \& {Amati}, L. 2013, \apj, 777,
  132

\bibitem[{{Fender} {et~al.}(2004){Fender}, {Belloni}, \& {Gallo}}]{Fender2004}
{Fender}, R.~P., {Belloni}, T.~M., \& {Gallo}, E. 2004, \mnras, 355, 1105

\bibitem[{{Fern{\'a}ndez} {et~al.}(2019){Fern{\'a}ndez}, {Tchekhovskoy},
  {Quataert}, {Foucart}, \& {Kasen}}]{Fernandez2019}
{Fern{\'a}ndez}, R., {Tchekhovskoy}, A., {Quataert}, E., {Foucart}, F., \&
  {Kasen}, D. 2019, \mnras, 482, 3373

\bibitem[{{Fishbone} \& {Moncrief}(1976)}]{Fishbone_Moncrief_1976_ApJ}
{Fishbone}, L.~G. \& {Moncrief}, V. 1976, \apj, 207, 962

\bibitem[{{Gammie} {et~al.}(2003){Gammie}, {McKinney}, \&
  {T{\'o}th}}]{Gammie_2003}
{Gammie}, C.~F., {McKinney}, J.~C., \& {T{\'o}th}, G. 2003, \apj, 589, 444

\bibitem[{{Giommi} {et~al.}(2021){Giommi}, {Perri}, {Capalbi}, {D'Elia},
  {Barres de Almeida}, {Brandt}, {Pollock}, {Arneodo}, {Di Giovanni}, {Chang},
  {Civitarese}, {De Angelis}, {Leto}, {Verrecchia}, {Ricard}, {Di Pippo},
  {Middei}, {Penacchioni}, {Ruffini}, {Sahakyan}, {Israyelyan}, \&
  {Turriziani}}]{Giommi2021}
{Giommi}, P., {Perri}, M., {Capalbi}, M., {et~al.} 2021, \mnras, 507, 5690

\bibitem[{{Globus} \& {Levinson}(2016)}]{GlobusLevinson2016}
{Globus}, N. \& {Levinson}, A. 2016, \mnras, 461, 2605

\bibitem[{{Godfrey} {et~al.}(2012){Godfrey}, {Lovell}, {Burke-Spolaor},
  {Ekers}, {Bicknell}, {Birkinshaw}, {Worrall}, {Jauncey}, {Schwartz},
  {Marshall}, {Gelbord}, {Perlman}, \& {Georganopoulos}}]{Godfrey2012}
{Godfrey}, L.~E.~H., {Lovell}, J.~E.~J., {Burke-Spolaor}, S., {et~al.} 2012,
  \apjl, 758, L27

\bibitem[{{Guidorzi} {et~al.}(2016){Guidorzi}, {Dichiara}, \&
  {Amati}}]{Guidorzi2016}
{Guidorzi}, C., {Dichiara}, S., \& {Amati}, L. 2016, \aap, 589, A98

\bibitem[{{Harris} \& {Krawczynski}(2006)}]{Harris2006}
{Harris}, D.~E. \& {Krawczynski}, H. 2006, \araa, 44, 463

\bibitem[{{Heger} {et~al.}(2005){Heger}, {Woosley}, \& {Spruit}}]{Heger2005}
{Heger}, A., {Woosley}, S.~E., \& {Spruit}, H.~C. 2005, \apj, 626, 350

\bibitem[{Hossein~Nouri {et~al.}(2018)Hossein~Nouri, Duez, Foucart, Deaton,
  Haas, Haddadi, Kidder, Ott, Pfeiffer, Scheel, \&
  Szilagyi}]{PhysRevD.97.083014}
Hossein~Nouri, F., Duez, M.~D., Foucart, F., {et~al.} 2018, Phys. Rev. D, 97,
  083014

\bibitem[{{Hubrig} {et~al.}(2020){Hubrig}, {J{\"a}rvinen}, {Sch{\"o}ller}, \&
  {Hummel}}]{Hubrig2020}
{Hubrig}, S., {J{\"a}rvinen}, S.~P., {Sch{\"o}ller}, M., \& {Hummel}, C.~A.
  2020, \mnras, 491, 281

\bibitem[{{Igumenshchev}(2008)}]{Igumenshchev2008}
{Igumenshchev}, I.~V. 2008, \apj, 677, 317

\bibitem[{Janiuk(2017)}]{Janiuk2017}
Janiuk, A. 2017, \apj, 837, 39

\bibitem[{{Janiuk}(2019)}]{Janiuk2019}
{Janiuk}, A. 2019, \apj, 882, 163

\bibitem[{{Janiuk} {et~al.}(2021){Janiuk}, {James}, \& {Palit}}]{Janiuk2021}
{Janiuk}, A., {James}, B., \& {Palit}, I. 2021, \apj, 917, 102

\bibitem[{Janiuk {et~al.}(2013)Janiuk, Mioduszewski, \&
  Moscibrodzka}]{Janiuk2013}
Janiuk, A., Mioduszewski, P., \& Moscibrodzka, M. 2013, \apj, 776, 105

\bibitem[{Janiuk {et~al.}(2018)Janiuk, Sapountzis, Mortier, \&
  Janiuk}]{JSFI177}
Janiuk, A., Sapountzis, K., Mortier, J., \& Janiuk, I. 2018, Supercomputing
  Frontiers and Innovations, 5

\bibitem[{{Janiuk} {et~al.}(2018){Janiuk}, {Sukova}, \& {Palit}}]{Janiuk2018}
{Janiuk}, A., {Sukova}, P., \& {Palit}, I. 2018, \apj, 868, 68

\bibitem[{{Janiuk} \& {Yuan}(2010)}]{JaniukYuan2010}
{Janiuk}, A. \& {Yuan}, Y.~F. 2010, \aap, 509, A55

\bibitem[{{Kathirgamaraju} {et~al.}(2019){Kathirgamaraju}, {Tchekhovskoy},
  {Giannios}, \& {Barniol Duran}}]{Kati2019}
{Kathirgamaraju}, A., {Tchekhovskoy}, A., {Giannios}, D., \& {Barniol Duran},
  R. 2019, \mnras, 484, L98

\bibitem[{{Kobayashi} {et~al.}(1997){Kobayashi}, {Piran}, \&
  {Sari}}]{Kobayashi1997}
{Kobayashi}, S., {Piran}, T., \& {Sari}, R. 1997, \apj, 490, 92

\bibitem[{{Krolik}(1999)}]{Krolik1999}
{Krolik}, J.~H. 1999, {Active galactic nuclei : from the central black hole to
  the galactic environment}

\bibitem[{{Kuan} {et~al.}(2021){Kuan}, {Suvorov}, \& {Kokkotas}}]{Kuan2021}
{Kuan}, H.-J., {Suvorov}, A.~G., \& {Kokkotas}, K.~D. 2021, \mnras, 508, 1732

\bibitem[{{Kubota} \& {Done}(2019)}]{2019MNRAS.489..524K}
{Kubota}, A. \& {Done}, C. 2019, \mnras, 489, 524

\bibitem[{Lazzati {et~al.}(2017)Lazzati, Deich, Morsony, \&
  Workman}]{Lazzati2017}
Lazzati, D., Deich, A., Morsony, B.~J., \& Workman, J.~C. 2017, \mnras, 471,
  1652

\bibitem[{{Lazzati} {et~al.}(2021){Lazzati}, {Perna}, {Ciolfi}, {Giacomazzo},
  {L{\'o}pez-C{\'a}mara}, \& {Morsony}}]{Lazzati2021}
{Lazzati}, D., {Perna}, R., {Ciolfi}, R., {et~al.} 2021, \apjl, 918, L6

\bibitem[{{Lee} \& {Ramirez-Ruiz}(2007)}]{Lee2007}
{Lee}, W.~H. \& {Ramirez-Ruiz}, E. 2007, New Journal of Physics, 9, 17

\bibitem[{{Levinson} \& {Globus}(2017)}]{LevinsonGlobus2017}
{Levinson}, A. \& {Globus}, N. 2017, \mnras, 465, 1608

\bibitem[{{Liska} {et~al.}(2020){Liska}, {Tchekhovskoy}, \&
  {Quataert}}]{Liska2020}
{Liska}, M., {Tchekhovskoy}, A., \& {Quataert}, E. 2020, \mnras, 494, 3656

\bibitem[{Liu {et~al.}(2015)Liu, Hou, Xue, \& Gu}]{Liu2015}
Liu, T., Hou, S.-J., Xue, L., \& Gu, W.-M. 2015, \apjs, 218, 12

\bibitem[{{Lloyd-Ronning} {et~al.}(2018){Lloyd-Ronning}, {Lei}, \&
  {Xie}}]{LloydRonning2018}
{Lloyd-Ronning}, N., {Lei}, W.-h., \& {Xie}, W. 2018, \mnras, 478, 3525

\bibitem[{{MacLachlan} {et~al.}(2013){MacLachlan}, {Shenoy}, {Sonbas}, {Dhuga},
  {Cobb}, {Ukwatta}, {Morris}, {Eskandarian}, {Maximon}, \&
  {Parke}}]{McLachlan2013}
{MacLachlan}, G.~A., {Shenoy}, A., {Sonbas}, E., {et~al.} 2013, \mnras, 432,
  857

\bibitem[{{McKinney} \& {Gammie}(2004)}]{McKinney2004}
{McKinney}, J.~C. \& {Gammie}, C.~F. 2004, \apj, 611, 977

\bibitem[{{McKinney} {et~al.}(2012){McKinney}, {Tchekhovskoy}, \&
  {Blandford}}]{McKinney_2012}
{McKinney}, J.~C., {Tchekhovskoy}, A., \& {Blandford}, R.~D. 2012, \mnras, 423,
  3083

\bibitem[{{McKinney} {et~al.}(2014){McKinney}, {Tchekhovskoy}, {Sadowski}, \&
  {Narayan}}]{2014MNRAS.441.3177M}
{McKinney}, J.~C., {Tchekhovskoy}, A., {Sadowski}, A., \& {Narayan}, R. 2014,
  \mnras, 441, 3177

\bibitem[{{Melia} {et~al.}(2001){Melia}, {Bromley}, {Liu}, \&
  {Walker}}]{2001ApJ...554L..37M}
{Melia}, F., {Bromley}, B.~C., {Liu}, S., \& {Walker}, C.~K. 2001, \apjl, 554,
  L37

\bibitem[{{Mizuta} {et~al.}(2018){Mizuta}, {Ebisuzaki}, {Tajima}, \&
  {Nagataki}}]{Mizuta2018}
{Mizuta}, A., {Ebisuzaki}, T., {Tajima}, T., \& {Nagataki}, S. 2018, \mnras,
  479, 2534

\bibitem[{Mochkovitch {et~al.}(1993)Mochkovitch, Hernanz, Isern, \&
  Martin}]{Mochkovitch_et_al_1993}
Mochkovitch, R., Hernanz, M., Isern, J., \& Martin, X. 1993, \nat, 361, 236

\bibitem[{{Moderski} {et~al.}(1998){Moderski}, {Sikora}, \&
  {Lasota}}]{moderski}
{Moderski}, R., {Sikora}, M., \& {Lasota}, J.~P. 1998, \mnras, 301, 142

\bibitem[{{Morales Teixeira} {et~al.}(2018){Morales Teixeira}, {Avara}, \&
  {McKinney}}]{2018MNRAS.480.3547M}
{Morales Teixeira}, D., {Avara}, M.~J., \& {McKinney}, J.~C. 2018, \mnras, 480,
  3547

\bibitem[{{Morsony} {et~al.}(2010){Morsony}, {Lazzati}, \&
  {Begelman}}]{Morsony}
{Morsony}, B.~J., {Lazzati}, D., \& {Begelman}, M.~C. 2010, \apj, 723, 267

\bibitem[{{Mo{\'s}cibrodzka} {et~al.}(2021){Mo{\'s}cibrodzka}, {Janiuk}, \& {De
  Laurentis}}]{ipole2021}
{Mo{\'s}cibrodzka}, M., {Janiuk}, A., \& {De Laurentis}, M. 2021, \mnras, 508,
  4282

\bibitem[{{Murguia-Berthier} {et~al.}(2021){Murguia-Berthier}, {Ramirez-Ruiz},
  {De Colle}, {Janiuk}, {Rosswog}, \& {Lee}}]{Murguia_Berthier2021}
{Murguia-Berthier}, A., {Ramirez-Ruiz}, E., {De Colle}, F., {et~al.} 2021,
  \apj, 908, 152

\bibitem[{{Narayan} {et~al.}(2012){Narayan}, {S{\"A} dowski}, {Penna}, \&
  {Kulkarni}}]{2012MNRAS.426.3241N}
{Narayan}, R., {S{\"A} dowski}, A., {Penna}, R.~F., \& {Kulkarni}, A.~K. 2012,
  \mnras, 426, 3241

\bibitem[{{Neilsen} {et~al.}(2013){Neilsen}, {Nowak}, {Gammie}, {Dexter},
  {Markoff}, {Haggard}, {Nayakshin}, {Wang}, {Grosso}, {Porquet}, {Tomsick},
  {Degenaar}, {Fragile}, {Houck}, {Wijnands}, {Miller}, \&
  {Baganoff}}]{neilsen2013}
{Neilsen}, J., {Nowak}, M.~A., {Gammie}, C., {et~al.} 2013, \apj, 774, 42

\bibitem[{{Noble} {et~al.}(2006){Noble}, {Gammie}, {McKinney}, \& {Del
  Zanna}}]{Noble_et_all_2006}
{Noble}, S.~C., {Gammie}, C.~F., {McKinney}, J.~C., \& {Del Zanna}, L. 2006,
  \apj, 641, 626

\bibitem[{{Obergaulinger} \& {Aloy}(2022)}]{obergaulinger}
{Obergaulinger}, M. \& {Aloy}, M.~{\'A}. 2022, \mnras, 512, 2489

\bibitem[{{Palit} {et~al.}(2019){Palit}, {Janiuk}, \&
  {Sukova}}]{Palit2019MNRAS}
{Palit}, I., {Janiuk}, A., \& {Sukova}, P. 2019, \mnras, 487, 755

\bibitem[{Paschalidis {et~al.}(2015)Paschalidis, Ruiz, \&
  Shapiro}]{Paschalidis_Ruiz_Shapiro_2015}
Paschalidis, V., Ruiz, M., \& Shapiro, S.~L. 2015, \apjl, 806, L14

\bibitem[{{Ponce} {et~al.}(2014){Ponce}, {Palenzuela}, {Lehner}, \&
  {Liebling}}]{Ponce2014}
{Ponce}, M., {Palenzuela}, C., {Lehner}, L., \& {Liebling}, S.~L. 2014, \prd,
  90, 044007

\bibitem[{{Ponti} {et~al.}(2015){Ponti}, {De Marco}, {Morris}, {Merloni},
  {Mu{\~n}oz-Darias}, {Clavel}, {Haggard}, {Zhang}, {Nandra}, {Gillessen},
  {Mori}, {Neilsen}, {Rea}, {Degenaar}, {Terrier}, \& {Goldwurm}}]{ponti2015}
{Ponti}, G., {De Marco}, B., {Morris}, M.~R., {et~al.} 2015, \mnras, 454, 1525

\bibitem[{{Prieto} {et~al.}(2016){Prieto}, {Fern{\'a}ndez-Ontiveros},
  {Markoff}, {Espada}, \& {Gonz{\'a}lez-Mart{\'\i}n}}]{2016MNRAS.457.3801P}
{Prieto}, M.~A., {Fern{\'a}ndez-Ontiveros}, J.~A., {Markoff}, S., {Espada}, D.,
  \& {Gonz{\'a}lez-Mart{\'\i}n}, O. 2016, \mnras, 457, 3801

\bibitem[{{Punsly}(2007)}]{Punsly2007}
{Punsly}, B. 2007, \mnras, 374, L10

\bibitem[{{Quataert} {et~al.}(1999){Quataert}, {Narayan}, \&
  {Reid}}]{1999ApJ...517L.101Q}
{Quataert}, E., {Narayan}, R., \& {Reid}, M.~J. 1999, \apjl, 517, L101

\bibitem[{{Reichart} {et~al.}(2001){Reichart}, {Lamb}, {Fenimore},
  {Ramirez-Ruiz}, {Cline}, \& {Hurley}}]{Reichart2001}
{Reichart}, D.~E., {Lamb}, D.~Q., {Fenimore}, E.~E., {et~al.} 2001, \apj, 552,
  57

\bibitem[{Rhoads(1999)}]{Rhoads1999}
Rhoads, J.~E. 1999, \apj, 525, 737

\bibitem[{{Ripperda} {et~al.}(2022){Ripperda}, {Liska}, {Chatterjee}, {Musoke},
  {Philippov}, {Markoff}, {Tchekhovskoy}, \& {Younsi}}]{Ripperdaetal}
{Ripperda}, B., {Liska}, M., {Chatterjee}, K., {et~al.} 2022, \apjl, 924, L32

\bibitem[{{Ruiz} {et~al.}(2020){Ruiz}, {Tsokaros}, \& {Shapiro}}]{Ruiz2020}
{Ruiz}, M., {Tsokaros}, A., \& {Shapiro}, S.~L. 2020, \prd, 101, 064042

\bibitem[{{Rusinek} {et~al.}(2020){Rusinek}, {Sikora}, {Kozie{\l}-Wierzbowska},
  \& {Gupta}}]{Rusinek2020}
{Rusinek}, K., {Sikora}, M., {Kozie{\l}-Wierzbowska}, D., \& {Gupta}, M. 2020,
  \apj, 900, 125

\bibitem[{{Sapountzis} \& {Janiuk}(2019)}]{Sap2019ApJ}
{Sapountzis}, K. \& {Janiuk}, A. 2019, \apj, 873, 12

\bibitem[{Sari {et~al.}(1999)Sari, Piran, \& Halpern}]{Sarietal1999}
Sari, R., Piran, T., \& Halpern, J.~P. 1999, \apjl, 519, L17

\bibitem[{{Scepi} {et~al.}(2021){Scepi}, {Begelman}, \& {Dexter}}]{2021scepi}
{Scepi}, N., {Begelman}, M.~C., \& {Dexter}, J. 2021, \mnras, 502, L50

\bibitem[{{Shibata} {et~al.}(2021){Shibata}, {Fujibayashi}, \&
  {Sekiguchi}}]{Shibata2021}
{Shibata}, M., {Fujibayashi}, S., \& {Sekiguchi}, Y. 2021, \prd, 104, 063026

\bibitem[{{Siegel} \& {Metzger}(2018)}]{Siegel2018}
{Siegel}, D.~M. \& {Metzger}, B.~D. 2018, \apj, 858, 52

\bibitem[{{Sikora} \& {Begelman}(2013)}]{SikoraBegelman2013}
{Sikora}, M. \& {Begelman}, M.~C. 2013, \apjl, 764, L24

\bibitem[{{Sikora} {et~al.}(2013){Sikora}, {Stasi{\'n}ska},
  {Kozie{\l}-Wierzbowska}, {Madejski}, \& {Asari}}]{Sikora2013}
{Sikora}, M., {Stasi{\'n}ska}, G., {Kozie{\l}-Wierzbowska}, D., {Madejski},
  G.~M., \& {Asari}, N.~V. 2013, \apj, 765, 62

\bibitem[{{Tamburini} {et~al.}(2020){Tamburini}, {Thid{\'e}}, \& {Della
  Valle}}]{2020MNRAS.492L..22T}
{Tamburini}, F., {Thid{\'e}}, B., \& {Della Valle}, M. 2020, \mnras, 492, L22

\bibitem[{Tchekhovskoy \& Giannios(2015)}]{Tchekhovskoy_Giannios2015}
Tchekhovskoy, A. \& Giannios, D. 2015, \mnras, 447, 327

\bibitem[{{Tchekhovskoy} {et~al.}(2011){Tchekhovskoy}, {Narayan}, \&
  {McKinney}}]{Tchekhovskoy_2011}
{Tchekhovskoy}, A., {Narayan}, R., \& {McKinney}, J.~C. 2011, \mnras, 418, L79

\bibitem[{{Ukwatta} {et~al.}(2011){Ukwatta}, {Dhuga}, {Morris}, {MacLachlan},
  {Parke}, {Maximon}, {Eskandarian}, {Gehrels}, {Norris}, \&
  {Shenoy}}]{Ukwatta2011}
{Ukwatta}, T.~N., {Dhuga}, K.~S., {Morris}, D.~C., {et~al.} 2011, \mnras, 412,
  875

\bibitem[{{van Velzen} \& {Falcke}(2013)}]{Falcke2013}
{van Velzen}, S. \& {Falcke}, H. 2013, \aap, 557, L7

\bibitem[{{Witzel} {et~al.}(2018){Witzel}, {Martinez}, {Hora}, {Willner},
  {Morris}, {Gammie}, {Becklin}, {Ashby}, {Baganoff}, {Carey}, {Do}, {Fazio},
  {Ghez}, {Glaccum}, {Haggard}, {Herrero-Illana}, {Ingalls}, {Narayan}, \&
  {Smith}}]{Witzel2018}
{Witzel}, G., {Martinez}, G., {Hora}, J., {et~al.} 2018, \apj, 863, 15

\bibitem[{{Witzel} {et~al.}(2021){Witzel}, {Martinez}, {Willner}, {Becklin},
  {Boyce}, {Do}, {Eckart}, {Fazio}, {Ghez}, {Gurwell}, {Haggard},
  {Herrero-Illana}, {Hora}, {Li}, {Liu}, {Marchili}, {Morris}, {Smith},
  {Subroweit}, \& {Zensus}}]{Witzel2021}
{Witzel}, G., {Martinez}, G., {Willner}, S.~P., {et~al.} 2021, \apj, 917, 73

\bibitem[{{Wu} {et~al.}(2016){Wu}, {Zhang}, {Lei}, {Zou}, {Liang}, \&
  {Cao}}]{wu2016}
{Wu}, Q., {Zhang}, B., {Lei}, W.-H., {et~al.} 2016, \mnras, 455, L1

\end{thebibliography}

\end{document}